\documentclass[useAMS,usenatbib, usegraphicx]{mn2e}

\usepackage{subfigure}
\usepackage[none]{hyphenat}
\usepackage[fleqn]{amsmath}
\usepackage{color}
\usepackage[total={17.8cm,24.0cm},centering]{geometry}

\title[Planetary Evaporation]{Planetary evaporation by UV \& X-ray radiation: basic hydrodynamics}
\author[James E. Owen \& Alan P. Jackson]{James E. Owen$^{1}$\thanks{E-mail: jowen@cita.utoronto.ca} and Alan P. Jackson$^{2}$\\
$^{1}$Canadian Institute for Theoretical Astrophysics, 60 St. George Street, Toronto, M5S 3H8, Canada.\\
$^{2}$Institute of Astronomy, Madingley Road, Cambridge, CB3 0DS, England.}
\begin{document}

\pagerange{\pageref{firstpage}--\pageref{lastpage}} \pubyear{2012}

\maketitle

\label{firstpage}

\begin{abstract}
We consider the evaporation of close in planets by the star's intrinsic EUV and X-ray radiation. We calculate evaporation rates by solving the hydrodynamical problem for planetary evaporation including heating from both X-ray and EUV radiation. We show that most close-in planets ($a<0.1$ AU) are evaporating hydrodynamically, with the evaporation occurring in two distinct regimes: X-ray driven, in which the X-ray heated flow contains a sonic point, and EUV driven, in which the X-ray region is entirely sub-sonic. The mass-loss rates scale as $L_X/a^2$ for X-ray driven evaporation, and as $\Phi_*^{1/2}/a$ for EUV driven evaporation at early times, with mass-loss rates of order $10^{10}-10^{14}$~g~s$^{-1}$. No exact scaling exists for the mass-loss rate with planet mass and planet radius, however, in general evaporation proceeds more rapidly for planets with lower densities and higher masses. Furthermore, we find that in general the transition from X-ray driven to EUV driven evaporation occurs at lower X-ray luminosities for planets closer to their parent stars and for planets with lower densities. 

Coupling our evaporation models to the evolution of the high energy radiation - which falls with time - we are able to follow the evolution of evaporating planets. We find that most planets start off evaporating in the X-ray driven regime, but switch to EUV driven once the X-ray luminosity falls below a critical value. The evolution models suggest that while `hot Jupiters' are evaporating, they are not evaporating at a rate sufficient to remove the entire gaseous envelope on Gyr time-scales. However, we do find that close in Neptune mass planets are more susceptible to complete evaporation of their envelopes. Thus we conclude that planetary evaporation is more important for lower mass planets, particularly those in the `hot Neptune'/`super Earth' regime.
\end{abstract}


\section{Introduction}
\label{introduction}
The majority of planets detected to date are found to be close ($<0.5$AU) to their parent star. At these small separations, and particularly at early times, they will be strongly irradiated.  The bolometric flux of the host star may inflate the planet's radius, above that which would be expected at larger separations, due to heating (e.g \citealt{baraffe2008}).  However it is the high energy radiation (UV and X-rays) that will be important for the evolution of the planet's upper atmosphere, where the gas temperatures may become close to the escape temperature. It is important to understand the process of planetary evaporation as it may have important evolutionary consequences for close-in planets, and it has been suggested that gas giants might entirely loose their gaseous envelopes through such a process, leaving behind a rocky core (e.g. \citealt{bjackson2010}). Observational evidence for planetary evaporation exists through the detection of extended atmospheres in several stellar lines that give planetary radii considerably larger than in the optical and infrared.  In particular the atmosphere extends beyond the Roche-radius, indicating the gas is no longer gravitationally bound to the planet.  The two best studied examples thus far are HD~209458b (\citealt{vidalmadjar2003, vidalmadjar2004}) and HD~188733b (\citealt{lecavelier2010}), which both have estimated mass-loss rates in the range $\sim10^{10-13}$ g s$^{-1}$.

Currently, theoretical models of planetary evaporation are limited and are derived under simplistic assumptions, additionally there is little consensus whether it is driven by Extreme-UV (EUV) or X-ray heating.  There is even debate as to whether evaporation is in the hydrodynamic limit, or proceeds through ballistic loss of particles (e.g. Jean's escape) where the upper atmosphere is no longer pressure dominated. {Perhaps the simplest approach is to assume that every photon received at the planet's surface is turned into mass-loss at some efficiency (\citealt{watson1981, lecavelier2007}). Such an approach can only provide an order of magnitude estimate of the mass-loss rate, though it is likely to be fairly accurate in the case of `energy-limited' evaporation, by which we mean that the dominant energy loss process is $P{\rm d} V$ work as discussed by \citet{watson1981}.  This may indeed represent an observable region of parameter space at late times in the case of EUV driven evaporation (\citealt{murrayclay2009}). Nonetheless such an approach masks many of the complexities associated with thermally driven evaporative winds. Thermally driven winds are free to absorb energy up to their sonic surface, which may in principle, be located far from the launch point (\citealt{parker1960}). This results in the interception of a much higher fraction of the stellar high-energy flux than the simple planetary disk, an effect \citet{lammer2003} discussed by introducing a further efficiency factor into the `energy-limited' formalism in the form of an expansion radius ($\beta$) based on the work of \citealt{watson1981}. The inclusion of the expansion radius has been dropped in recent years as  detailed models of EUV heating in hot Jupiters suggested very little energy is absorbed high in the flow, with most energy deposited near or at the base of the flow (e.g. Yelle et al. 2004; Garcia Munoz 2007; \citealt{leitzinger2011}). Although it is unclear whether EUV heated flows may absorb significant amount of energy high in the flow for planets with different properties compared to hot Jupiters, and whether the X-rays can provide significant heating to the gas at large atmospheric heights, as suggested by \citet{cecchi2006}.}  

{In addition, one must account for the reduction in the gravitational binding of the planetary atmosphere induced by the stellar gravitational field, as discussed by \citet{erkaev2007}.  In the energy-limited formalism this is accomplished by the introduction of another efficiency factor based on the assumption one only need overcome the Roche potential rather than the gravitational field of the planet in isolation\footnote{Note: later we discuss how the effect of the star's gravity appears through the derivative of the potential rather than its absolute value}.  Thus the energy limited formalism presented by \citet{erkaev2007} gives rise to a mass-loss rate of the form
\begin{equation}
\dot{m}=\eta\frac{L_{\rm HE}R_p^3}{4GM_p a^2K(R_{\rm Roche}/R_p)},
\label{eqn:elimited}
\end{equation} 
where $\eta$ is the `efficiency' of the flow,  $L_{\rm HE}$ is the high energy luminosity of the central star, $M_P$, $R_p$ \& $a$ are the planet mass, radius and orbital distance and $K(R_{\rm{Roche}}/R_P)$ accounts for the reduction in the planetary binding energy due to the Roche lobe. Eq. \ref{eqn:elimited} has been used for several parameter studies of both EUV evaporation (e.g. Lammer et al. 2009) and X-ray evaporation (e.g. \citealt{davis2009,jackson2012}); however,  the `energy limited' approach possesses tunable parameters, for which there is little reason to expect to be independent of the physical properties of the system.  As such Eq~\ref{eqn:elimited} is of limited use in understanding the true details of planetary evaporation over a range of parameter space. } 

Further, there is no reason that evaporation must occur in the energy-limited regime and it may instead be that radiative losses dominate the energy budget rather than $P{\rm d} V$ work. In the case of EUV evaporation at high luminosities \citet{murrayclay2009} showed that this results in a qualitatively different scaling with both the high-energy luminosity and orbital distance.  It is also necessary to determine whether the planetary evaporation is indeed occurring in the hydrodynamic limit, for which one needs full solutions to verify that the flow is pressure dominated all the way to the sonic surface. 

{Several hydrodynamic studies of planetary evaporation have been undertaken that go beyond the `energy-limited' formalism. \citet{yelle2004}, \citet{tian2005} and \citet{murrayclay2009} calculate the hydrodynamic escape for a pure EUV heated model, while \citet{yelle2004} additionally solves for the hydrogen and helium structure of the flow.  Of these only \citet{murrayclay2009} solve for the EUV radiation transfer and heating rate directly. However, both \citet{yelle2004} and \citet{tian2005} use fixed heating efficiencies (for photo-electric heating in the case of \citealt{yelle2004} and the total heating efficiency in the case of \citealt{tian2005}).  \citet{penz2008} include both X-ray and EUV, but they set the X-ray heating rate to a fixed efficiency, which is then varied between 10 and 60 per cent to obtain various flow solutions.  \citet{cecchi2006} consider pure X-ray heating of a static plane-parallel atmosphere and determine the X-ray driven heating with consideration of the hydrogen and helium photochemistry.  They conclude that at high X-ray luminosities similar to those found around young stars (e.g. $L_X\sim10^{30}$ erg s$^{-1}$), it may be X-rays that dominate the heating.  Conversely several hydrodynamical studies of EUV evaporation have been successfully applied to the observations of HD 209458b, correctly reproducing the observed mass-loss rates of $\sim 10^{10}-10^{11}$ g s$^{-1}$, suggesting X-rays may be unimportant in this particular case (e.g. \citealt{garcia2007,murrayclay2009,koskinen2010a}).  What is not clear, however, is the structure of X-ray driven flows, the interactions that take place between a hydrodynamic flow that has X-ray and EUV heated regions and how a flow may transitions from X-ray dominated to EUV dominated. }

As a prelude to performing multi-dimensional hydrodynamic calculations, in this work we build 1D, on-axis, hydrodynamical solutions to the problem of planetary evaporation including both EUV and X-ray radiation fields under the approximation of radiative equilibrium.  This allows us to understand whether planetary evaporation will be `energy-limited' or not, whether it will occur in the hydrodynamic limit, and what the interactions between the EUV and X-ray fields yield in terms of evaporation. Our paper is organised as follows: in Section~\ref{sec:heat} we discuss the various heating mechanisms; in Section~\ref{sec:basichydro} we derive our hydrodynamic solutions for EUV and X-ray heating; in Section~\ref{sec:evol} we consider simple evolution of planets undergoing evaporation; in Section~\ref{sec:discuss} we discuss the implications of our results and our assumptions, and finally we present our conclusions in Section~\ref{sec:conc}. 

\section{Heating Mechanism}
\label{sec:heat}
Solar-type stars are known to produce large amounts of high-energy radiation in the form of UV and X-ray photons. These high-energy photons originate in the stellar corona, and at early times may have a luminosity $\sim 10^{-3}$ of the bolometric luminosity (e.g. \citealt{guedel2004}). Furthermore, this high-energy radiation falls off with age (e.g. \citealt{jackson2012}) in association with stellar spin-down (e.g. \citealt{hempelmann1995, guedel2004, wright2011}). Therefore, the impact the high-energy photons have on the planetary atmosphere will also vary with time. 

Young pre-main sequence stars also emit large Far-UV (FUV) fluxes (e.g. \citealt{ingleby2011}), that could be responsible for driving efficient evaporation from disks (e.g. \citealt{gorti2011}). However, the main FUV heating mechanism to reach the high temperatures ($\ga 1000$K) necessary for planetary evaporation is photoelectric heating from dust grains (e.g. \citealt{tielens1985a, gorti2004}). Due to the high dust-to-gas mass ratios required for efficient FUV heating we will at this stage neglect the influence of FUV radiation, though we note that FUV heating at lower temperatures may have some influence through other heating channels such as H$_2$ pumping. Thus while it is unlikely that the FUV can be a dominant driving mechanism it may induce slight differences in the pure X-ray/EUV evaporation discussed here.

\subsection{X-rays}
\label{sec:Xrayheat}
X-ray photons span a large range in energy $h\nu \approx 0.1-10$~keV, with the `soft' X-rays ($h\nu\la1-2$~keV) responsible for heating in the flux range expected for close-in planets. { \citet{cecchi2006} considered X-ray heating for a plane-parallel planetary atmosphere with similar fluxes to those found around close-in planets, and concluded that photo-electric heating could heat planetary atmospheres to sufficiently high temperatures to allow evaporation.}  For observed X-ray luminosities, and gas densities in planetary atmospheres, heating typically occurs to neutral column densities of $N=10^{22}$ cm$^{-2}$, with the gas reaching temperatures in the range of a few hundred K to $\sim 10^4$~K (e.g. \citealt{glassgold2004, cecchi2006, ercolano2009}). The dominant heating mechanism is photo-electron generation by the K-shells of metals, with Oxygen and Carbon being the most important. Furthermore, the heating time-scale will always be the quickest time-scale in the problem (simply the thermalisation time-scale of the hot, $\sim 10^6$~K, photoelectrons).

\begin{figure}
\centering
\includegraphics[width=\columnwidth]{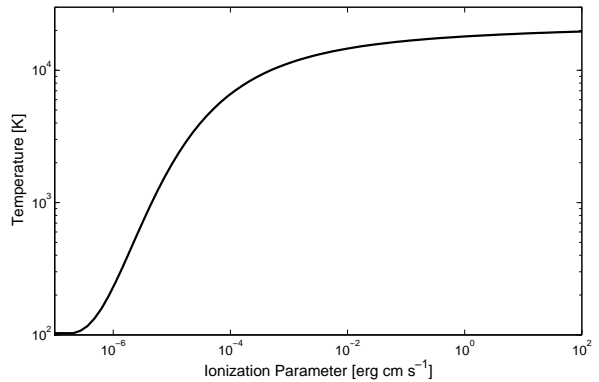}
\caption{Temperature-ionization parameter relation from \citet{owen2010}.}\label{fig:Txi}
\end{figure}

Cooling in X-ray heated gas can occur through several processes, though for the expected range of temperatures here line cooling will dominate, as X-ray recombination and thermal bremsstrahlung are only important at much higher luminosities. Line cooling is also dominated by metals, and again Carbon and Oxygen are the most important. In radiative equilibrium the temperature of X-ray heated gas can be described as a monotonic function of the ionization parameter - $\xi=L_X/na^2$ - (e.g \citealt{tarter1969,igea1999, owen2010}), where $L_X$ is the X-ray luminosity, $n$ the number density and $a$ the distance to the X-ray source. Such temperature-ionization parameter ($T=f(\xi)$) relations were calculated in \citet{owen2010} for use in discussion of disk photoevaporation around solar-type stars. Given the similar ionization parameters expected, since the escape temperature from a planetary atmosphere is similar to the escape temperature from the inner disk (1-10 AU) around a solar-type star, we will make use of the temperature-ionization parameter relation of \citet{owen2010} in this work, as shown in Fig.~\ref{fig:Txi}.

Furthermore, given that the same metals dominate both heating and cooling of the gas we may include a basic metallicity scaling. To do this we make use of the results of \citet{ercolano2010} for the heating of protoplanetary disks. Interpreting their calculated scaling of the mass-loss rate with metallicity ($\dot{M}_w\propto Z^{-0.77}$) within the theoretical framework presented in \citet{owen2012}, it is easy to show that the ionization parameter scales with metallicity as $\xi\propto Z^{-0.77}$. Thus, for the simple evaporation models we are developing in this work, we incorporate metallicity effects using the above scaling.

\subsection{Extreme-UV}
\label{sec:euvheat}
Extreme-UV (EUV) photons have sufficient energy to directly ionize hydrogen atoms ($h\nu\ga 13.6$ eV), resulting in fully ionized gas at $\sim 10^{4}$~K. In radiative equilibrium the  the ionized gas readily produces further EUV photons by recombination, so the gas thermostats to $10^{4}$~K, even in the presence of mild heating/cooling from other radiation (e.g. X-rays), thus we use an isothermal equation of state for EUV heated gas. The extent of the EUV heated region can then be found by balancing recombination with ionization. 

\citet{murrayclay2009} determined that at low EUV luminosities typical of stars with Gyr ages, radiative cooling is no longer the dominant cooling mechanism in an evaporative flow, and instead cooling through $P{\rm d} V$ work dominates (as much as $80\%$ of the energy went into $P{\rm d}V$ work). At this stage we are interested at looking at times when EUV evaporation can contribute significantly to the mass-loss from the planet, which will happen at early times when the EUV luminosity is high enough to be in radiative equilibrium. However, we do take into account this `energy-limited' EUV evaporation when we consider the evolution of example planets in Section~\ref{sec:evol}. 

\section{Basic Hydrodynamics}
\label{sec:basichydro}
In this section we discuss the basics of an evaporative hydrodynamic wind from a planet. To do this we solve the steady-state hydrodynamic problem along a streamline connecting planet and star in a frame co-rotating with the planet  using similar model setups to previous hydrodynamic models (e.g. \citealt{tian2005,penz2008,murrayclay2009}), although we do not assume a fixed heating efficiency for the X-rays are rather calculated the gas temperature directly. As in \citet{murrayclay2009} we neglect the Coriolis force, which can be considered as small in the trans-sonic region of the flow as $|\Omega||u|\ll$ gravity/pressure. Therefore, the governing equations are the steady-state continuity and momentum equations
\begin{eqnarray}
\frac{\partial}{\partial r}\left(\rho u r^2\right) & = & 0,\label{eqn:cont} \\
u\frac{\partial u}{\partial r} & = & -\frac{1}{\rho}\frac{\partial P}{\partial r}-\frac{\partial \Psi_{\rm eff}}{\partial r},\label{eqn:mom}
\end{eqnarray}
where $r$ is the distance from the centre of the planet, $\rho$ is the density of the flow, $u$ is the flow velocity, $P$ is the pressure and $\Psi_{\rm eff}$ is the effective (Roche) potential in the co-rotating frame, given by
\begin{equation}
\label{eqn:psieff}
\begin{split}
\Psi_{\rm eff} &=-\frac{GM_p}{r}-\frac{GM_*}{|a-r|}\\
               &\quad -\frac{1}{2}\frac{G\left(M_p+M_*\right)}{a^3}\left[a\left(\frac{M_*}{M_*+M_p}\right)-r\right]^2.
\end{split}
\end{equation}

We adopt radiative equilibrium, which holds in the limit of large fluxes (most of the received energy is emitted as cooling radiation rather than lost to $P{\rm d}V$ work), an assumption we will justify in detail in Section~\ref{sec:assump}. In a region that is exposed to both EUV and X-ray radiation, we adopt an isothermal equation of state as discussed in Section~\ref{sec:euvheat}, given by
\begin{equation}
P_{\rm EUV}=\rho c_{\rm EUV}^2,
\end{equation}
where $c_{\rm EUV}$ is the isothermal sound speed in $\sim 10^4 K$ gas, for which we adopt a value of $10$~km~s$^{-1}$. In gas that is only exposed to X-rays, we adopt a temperature-ionization parameter relation, $T=f(\xi)$, where $f$ is a monotonic function taken from the previous X-ray calculations of \citet{owen2010, owen2011}, as disused in Section~\ref{sec:Xrayheat}\footnote{We note that this function is empirical and simply use the label $f$ to denote it in this work}.

\begin{figure*}
\centering
\subfigure[X-ray driven evaporation]{\includegraphics[width=0.8\columnwidth]{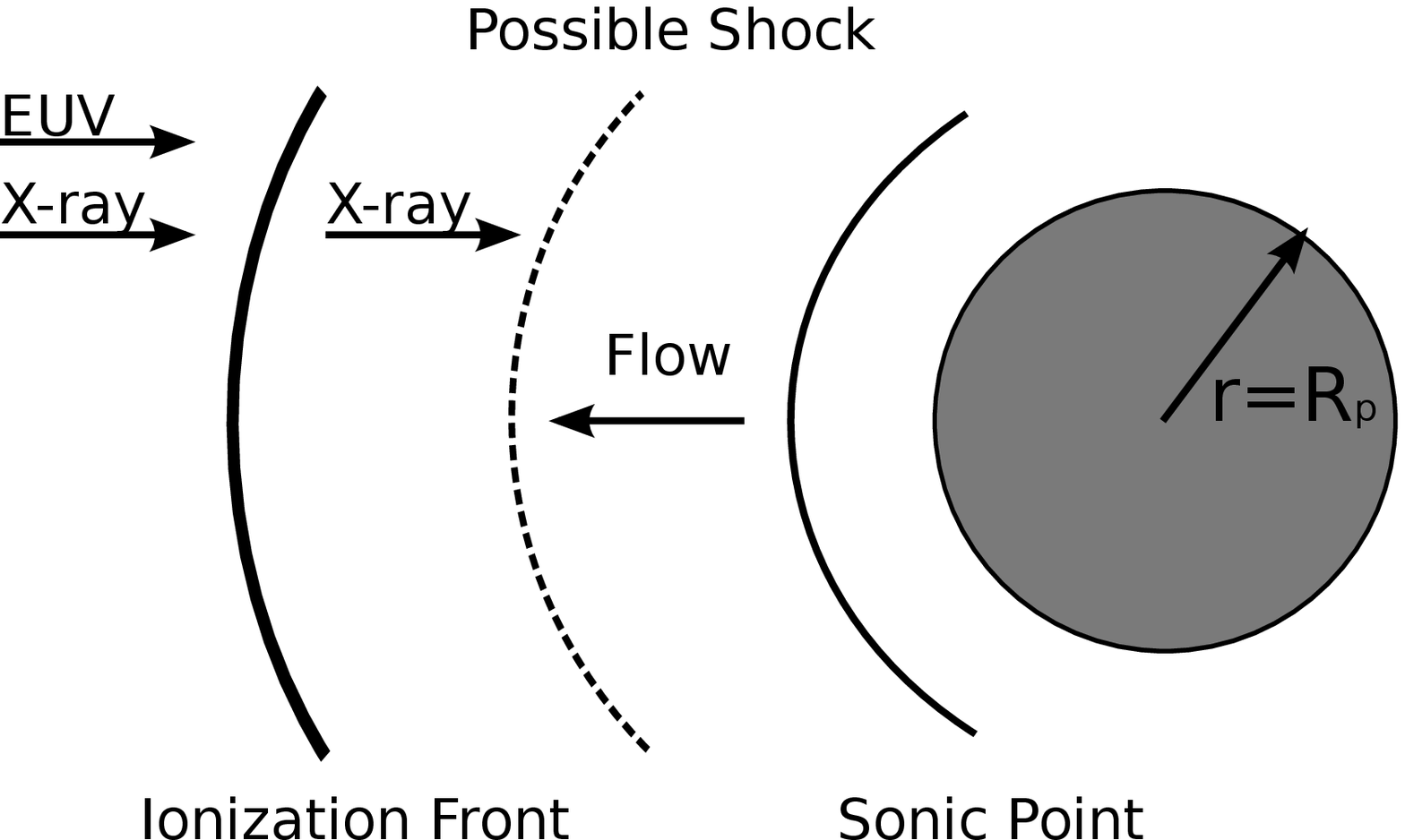}\label{sfig:Xray}}\hspace{0.09\textwidth}
\subfigure[EUV driven evaporation]{\includegraphics[width=0.8\columnwidth]{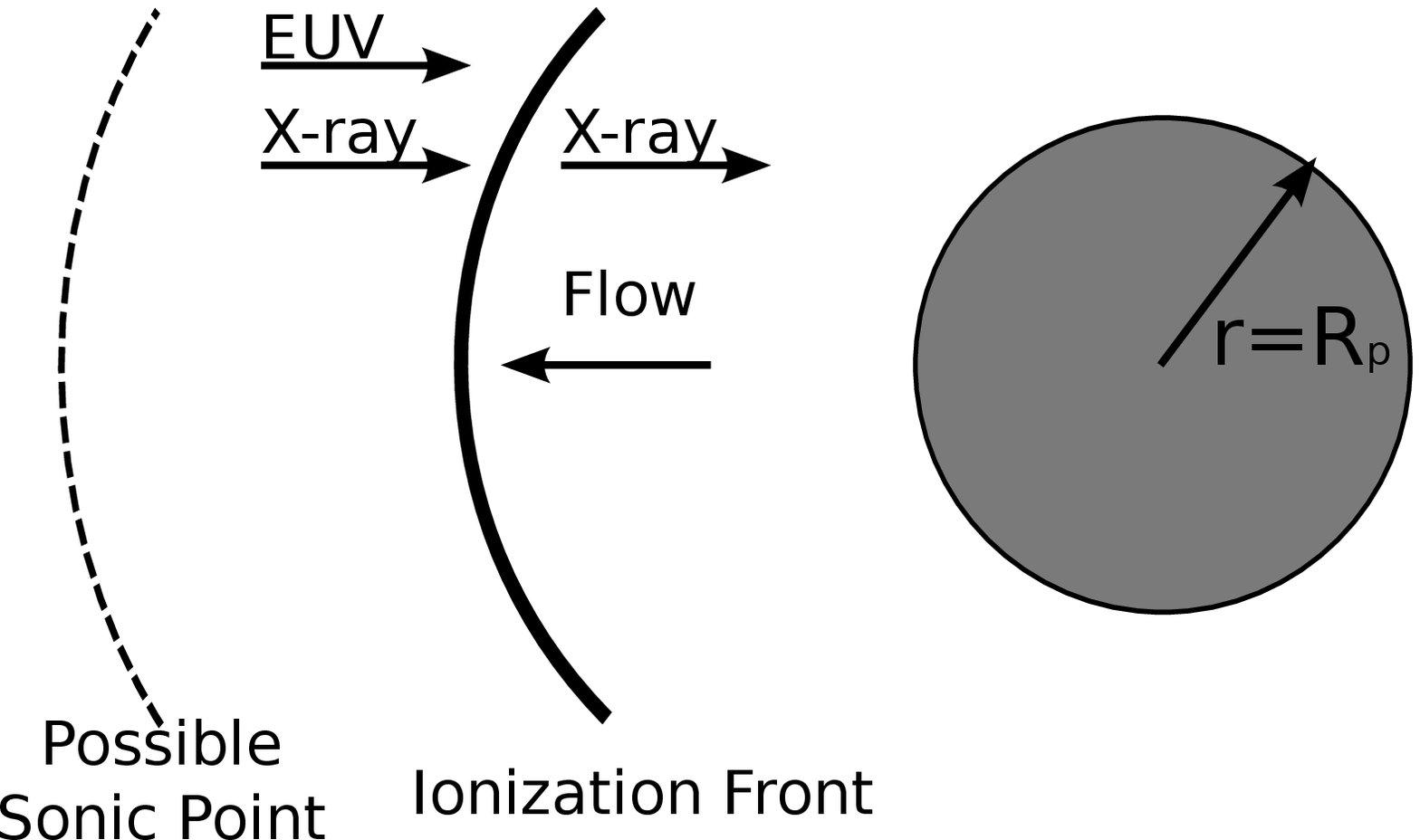}\label{sfig:EUV}}
\caption{Flow topologies for the X-ray driven and EUV driven evaporation cases. In the case of X-ray driven evaporation the X-ray heated flow passes through a sonic surface, then possibly shocks before passing through the ionization front and becoming EUV heated. In the case of EUV driven evaporation a sub-sonic X-ray flow passes through the ionization front, the EUV heated flow then is either supersonic or proceeds to pass through an EUV heated sonic point. See text for a discussion of whether the shock exists in the case of X-ray driven evaporation, and whether the sonic point exists in the case of EUV driven evaporation.}\label{fig:flowtop}
\end{figure*}

The transition from an EUV to X-ray heated wind will occur at an ionization front located at a distance $r=R_{\rm IF}$ from the planet. Such a flow is qualitatively similar to those described by \citet{johnstone1998} for the photoevaporation of disks illuminated by external UV radiation, where in \citet{johnstone1998} an ionization front separates an EUV heated flow from an FUV heated flow. The location of the ionization front relative to the sonic point is crucial in determining which heating mechanism sets the mass-loss rate (e.g. \citealt{johnstone1998, hollenbach2000, owen2010}), since the flow above the sonic point is causally disconnected from that below it. Therefore, in a similar manner to the FUV or EUV dominated photoevaporation of \citet{johnstone1998}, we can identify two separate cases of planetary evaporation: X-ray dominated, where the sonic transition occurs in the X-ray heated flow, and EUV dominated, where the ionization front occurs before any sonic transition in the X-ray heated region of the flow. In the latter case a partially neutral, sub-sonic, X-ray flow, passes through an ionization front into an ionized, transonic, EUV flow.  These two cases are illustrated schematically in Fig.~\ref{fig:flowtop}, where the left hand panel shows X-ray driven evaporation and the right hand panel EUV driven evaporation, where the EUV heated region is always closer to the central star as the X-rays have a much lower absorption cross-section.

Here, we are only interested in mass-loss rates, which do not depend on the structure of the flow beyond the sonic point, so we do not solve for the EUV/X-ray transition in the supersonic region of an X-ray dominated flow.  In reality the X-ray wind may shock and form another sub-sonic X-ray flow, which then passes through an ionization front into a supersonic EUV heated flow, or it may just supersonically pass through a rarefaction ionization front directly into a supersonic EUV heated flow.  The exact configuration (shock or no shock) will be determined by the Mach number upstream of the ionization front (\citealt{spitzer1978}).  In an EUV dominated flow, however, we do solve for the EUV/X-ray transition, as the X-ray heated flow will be entirely sub-sonic, and the associated possible sonic point will be in the EUV heated region, determined by the escape temperature at the ionization front.  The position of the ionization front is set by balancing recombination and ionization at the ionization front with the mass-flux ($\dot{m}_X$) passing through the ionization front.  That is
\begin{equation}\label{eqn:EUV1}
\frac{\Phi_*}{4\pi a^2}=A\frac{\alpha_r}{16 \pi^2}\left(\frac{\dot{m}_X}{\mu\, c_{\rm EUV}}\right)^2 R_{\rm IF}^{-3},
\end{equation}
where $\Phi_*$ is the luminosity of EUV photons (photons per second), $\alpha_r=2.6\times10^{-13}$ cm$^3$ s$^{-1}$ is the recombination coefficient for hydrogen at $10^4$~K, $\mu$ is the mean particle weight, and A is an order unity geometry factor that takes into account how steeply the density falls off in the ionized portion of the flow, with typically $A\approx 1/3$ (e.g \citealt{johnstone1998}). Therefore, since Eq.~\ref{eqn:EUV1} gives $R_{\rm IF}\propto \dot{m}_X^{2/3}$, there is a critical $\dot{m}_X$ at which the ionization front occurs at the same radius as the X-ray sonic surface ($R_{\rm IF}=R_s$). At an $\dot{m}_X$ below this critical value the flow will be EUV dominated, as the X-ray portion of the flow is entirely sub-sonic, whereas at an $\dot{m}_X$ above this critical value the flow will be X-ray dominated, as the sonic point occurs before $R_{\rm IF}$. As we shall show in Section~\ref{sec:X1}, the sonic surface for an X-ray dominated flow typically occurs within a few planetary radii of the planet, so if we characterise $R_s$ by $R_s=\beta R_p$, then $\beta$ is of order unity. Putting astrophysical numbers in Eq.~\ref{eqn:EUV1} we can determine the EUV luminosity necessary for the ionization front to occur before any sonic transition in the X-ray flow, and thus for which the flow will become EUV dominated as
\begin{equation}
\begin{split}
\Phi_* &\ge 10^{40} {\rm\, s}^{-1} \left(\frac{a}{0.1\rm { \,AU}}\right)^2\left(\frac{\dot{m}_X}{10^{12}\rm { \,g\;\,  s}^{-1}}\right)^2\left(\frac{A}{1/3}\right)\\
       &\quad \times \left(\frac{\beta}{1.5}\right)\left(\frac{R_p}{10 R_{\oplus}}\right)^{-3}.
\end{split}
\end{equation}
Given these values we can expect that there will be regions of parameter space corresponding to either EUV driven flows or X-ray driven flows as both can occur under sensible physical conditions. As we shall show later $\dot{m}_X\propto L_X/a^2$ and will scale with planet radius steeper than $R_p^{3/2}$, thus we can expect EUV dominated evaporation to occur for small planets, low X-ray luminosities and at large distances from the central star.

\subsection{X-ray driven}
\label{sec:X1}
If we assume the sonic point occurs before the ionization front, then we can proceed in calculating the transonic flow solution. As mentioned above, at this stage we are purely interested in obtaining mass-loss rates, so we ignore any interactions with the EUV in the super-sonic portions of the flow (we are essentially setting $\Phi_*=0$ in this sub-section). By making use of the fact $\xi=L_X\mu/\rho a^2$, and given $a$ is fixed for planets in approximately circular orbits, we see immediately that temperature is purely a function of density, or more precisely the flow is barotropic\footnote{The gas behaves as if it has a ratio of specific heats, \hbox{$\gamma=1-\frac{{\rm d}\log f}{{\rm d}\log\xi}$}.}. 

\subsubsection{Basic Structure}

\begin{figure*}
\centering
\includegraphics[width=0.85\textwidth]{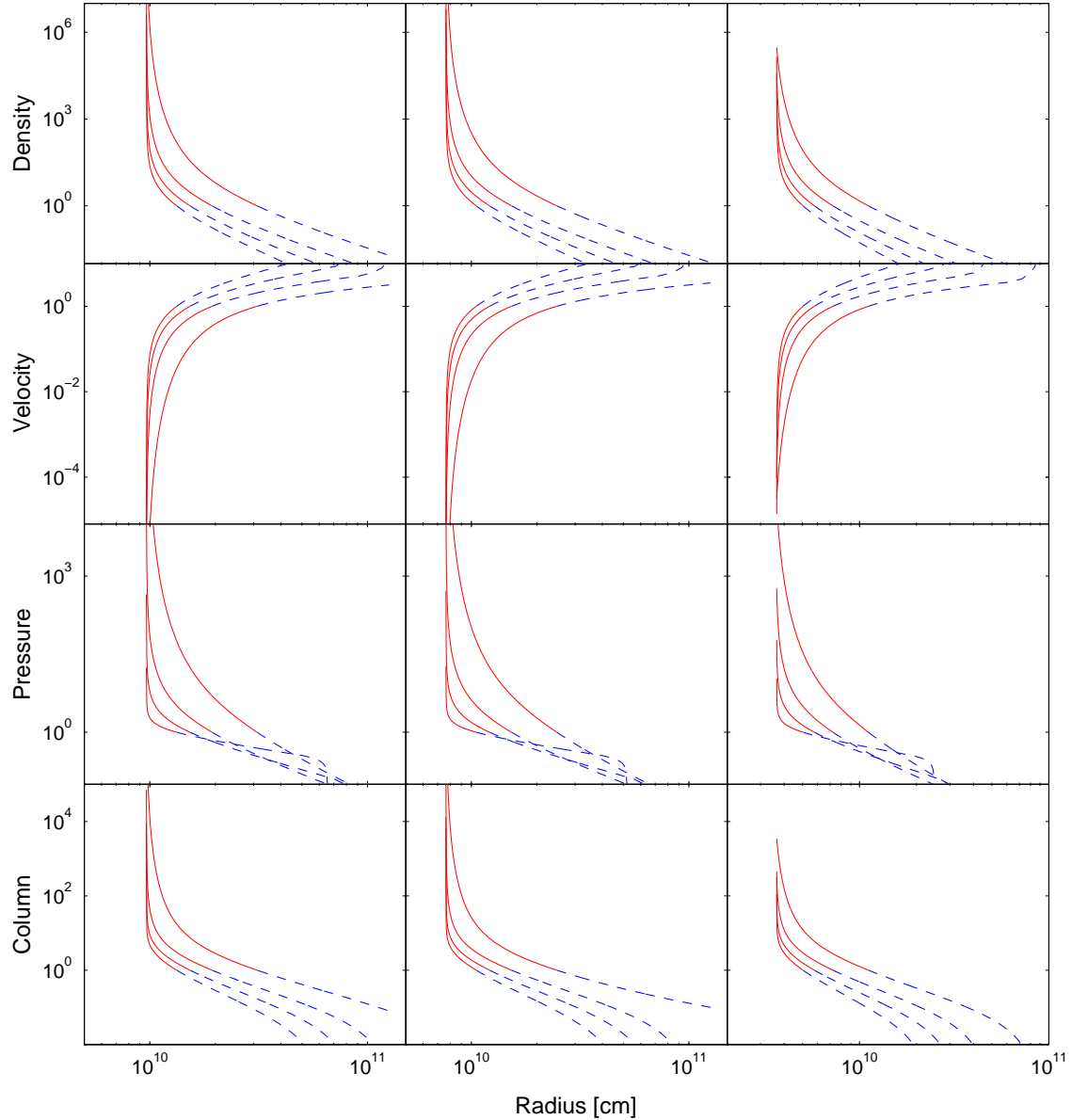}
\caption{Density, velocity, pressure and column density profiles for solutions to the X-ray evaporation problem, for various values of the temperature ionization-parameter slope ($\alpha$=0.2, 0.4, 0.6 \& 0.8).  All profiles are scaled to their value at the sonic surface. The solid line indicates the sub-sonic portion of the flow while the dashed line shows the super-sonic portion. The first column shows the solutions for a Jupiter mass planet with density 0.5~g~cm$^{-3}$, the second column a Jupiter mass planet with density 1~g~cm$^{-3}$, and the final column a Neptune mass planet with density 0.5~g~cm$^{-3}$.}\label{fig:flow_sim}
\end{figure*}

Before using the realistic temperature-ionization parameter function calculated by \citet{owen2010}, it is instructive to first set $f(\xi)\propto\xi^\alpha$ in order to analyse the basic properties of an X-ray driven flow. 
Using Eq.~\ref{eqn:cont} and the ideal gas equation we can re-write Eq.~\ref{eqn:mom} in the form
\begin{equation}
\label{eqn:crit1}
\begin{split}
\left[u^2-\left(1-\alpha\right)c_X^2\right]\frac{\partial \log u}{\partial r} & \\
& \hspace{-13ex} =\left(1-\alpha\right)c_X^2\left[\frac{2}{r}-\frac{1}{\left(1-\alpha\right)c_X^2}\frac{\partial \Psi_{\rm eff}}{\partial r}\right],
\end{split}
\end{equation}
where we define $c_X^2$ as an isothermal sound speed for a gas temperature $T$. Eq.~\ref{eqn:crit1} indicates that we can only obtain accelerating solutions ($\partial u/\partial r>0$) if $\alpha < 1$. This is trivial to understand, as one is unable to obtain a negative pressure gradient in an outflow if $\alpha\ge 1$ $(\partial \log P/\partial \log \rho < 0)$.  For $\alpha < 1$, Eq.~\ref{eqn:crit1} possesses a critical point at $u=\sqrt{1-\alpha}\,c_X$, which can be identified as the sonic point, such that to obtain a sub-sonic to super-sonic transition one requires that
\begin{equation}
\label{eqn:sonic1}
c_X^2=\frac{1}{1-\alpha}\frac{R_s}{2}\left.\frac{\partial\Psi_{\rm eff}}{\partial r} \right|_{r=R_s}.
\end{equation}
Such a condition has a very important effect - it fixes the temperature, and hence density, at the sonic surface.  In essence, since $\rho$ and $u$ are specified at this fixed radius, the mass-flux is also fixed here. 

Eq.~\ref{eqn:sonic1} also indicates that there is a maximum radius at which the sonic surface may occur, as one  requires $\partial \Psi_{\rm eff}/\partial r >0$.  The Roche-radius thus ultimately represents the maximum radius for a thermally driven outflow. Obviously planets with radii greater than the Roche-radius will be undergoing dynamical Roche-lobe overflow (e.g. \citealt{jaritz2005}), rather than wind driven mass-loss.  Furthermore, we can now calculate the entire flow solution by making use of the fact the Bernoulli function is conserved along a streamline in a barotropic flow (so numerical integrations of the fluid equations are not required).  Solving for the stream-function of a streamline connecting the planet and star we find
\begin{equation}
\label{eqn:bern1}
\begin{split}
\frac{u^2}{2} - \frac{1}{\alpha} \frac{R_s}{2} \left.\frac{\partial\Psi_{\rm eff}}{\partial r}\right|_{r=R_s} \left(\frac{\rho}{\rho_s}\right)^{-\alpha} + \Psi_{\rm eff}(r) & \\
& \hspace{-34ex} = \left(\frac{1}{2} - \frac{1}{\alpha}\right) \frac{R_s}{2} \left.\frac{\partial\Psi_{\rm eff}}{\partial r}\right|_{r=R_s} + \Psi_{\rm eff}(R_s),
\end{split}
\end{equation}
where the RHS of Eq.~\ref{eqn:bern1} has been evaluated at the sonic surface. Eq.~\ref{eqn:bern1} specifies a {\it family} of flow solutions which depend on the chosen location of the sonic-surface, $R_s$. In reality the chosen flow solution, and hence chosen $R_s$, is set by dynamical balance in the upper atmosphere of the planet, and where the atmosphere transitions from being X-ray heated to heated by the star's bolometric luminosity (which we identify as the planetary radius, $R_p$). In Fig.~\ref{fig:flow_sim} we show density, velocity, pressure and column density profiles for the solution to this simple problem, varying the planetary density, mass, and the index $\alpha$.  From this we can see that lower values of $\alpha$ result in broader profiles, as do lower planetary masses and densities.

One of the important features of flows with an effective ratio of heat capacities $<1$ is the steep gradients in the very sub-sonic portions of the flow, close to the planet. In these very sub-sonic regions we can solve for the density structure by neglecting the kinetic energy term in the flow (first term on the LHS of Eq.~\ref{eqn:bern1}), and assuming the effective potential is given by $\Phi=-GM_p/r$.  Solving for the density structure in the sub-sonic flow in this way we obtain
\begin{equation}\label{eqn:subsonic}
\left(\frac{\rho}{\rho_s}\right)=\left[1+\frac{3}{2}\alpha-2\alpha\left(\frac{R_s}{R}\right)\right]^{-1/\alpha}.
\end{equation}
The pressure scale height of this atmosphere can be compared to the scale height of a typical underlying bolometrically heated planetary atmosphere.  Expanding about $R_p$ it is easy to show
\begin{equation}
\frac{\partial \log P}{\partial \log r} \sim -(r-R_p)^{-1}.
\end{equation}
Since this diverges as one approaches $R_p$, it will always tend to a value considerably steeper than the typical scale height of the planetary atmosphere ($\ga 100$km). This implies that while the position of the sonic surface will be sensitive to the planetary radius, the flow solution will be \emph{highly insensitive} to the structure of the planetary atmosphere below.  For a given change in the penetration depth of the flow (e.g. due to a change in the planet's atmospheric structure, or the X-ray luminosity) $\Delta P_{\rm flow}\gg\Delta P_{\rm atm}$, so a large change in the structure of the planetary atmosphere causes a very small change in resultant flow solution.  Thus, the evaporation rates are sensitive to the intrinsic properties of planetary mass and radius, but very insensitive to the atmospheric structure at $R_p$.  As such, though the material composition of the atmosphere may be important through metallicity effects in the flow, we can neglect any details of the structure of the planetary atmosphere below the flow.

With this in mind, and taking $R_p$ to be the point at which density diverges towards infinity, we can estimate $R_s/R_p$ in terms of $\alpha$.  Setting $\rho\rightarrow\infty$ in Eq.~\ref{eqn:subsonic} we find
\begin{equation}
R_s=\left(\frac{3}{4}+\frac{1}{2\alpha}\right)R_p\label{eqn:RpRs}.
\end{equation}
Thus we see that unless $\alpha \ll 1$ the sonic surface generally occurs within a few planetary radii, and that X-ray heating cannot support flows with sonic surfaces at very large distances from the planet surface, unless the flow is close to isothermal. This is what will ultimately limit the attainable mass-loss rates, and it is unlikely that the planetary wind can have a sonic surface close to the Roche-radius. 

\subsubsection{Realistic Solutions}
The flow solutions discussed above are qualitatively useful in understanding the features of X-ray flow, but should not be taken as accurate, representative, models for the evaporation of planetary atmospheres. In order to calculate more representative solutions, we need to use a realistic temperature-ionization parameter relation to calculate the flow solutions, as discussed above.  In Appendix~A we show that it is possible to obtain a family of transcendental solutions for the ionization parameter in the flow for different choices of the sonic point, using the same techniques used to obtain Eq.~\ref{eqn:bern1}.

\begin{figure*}
\centering
\includegraphics[width=\textwidth]{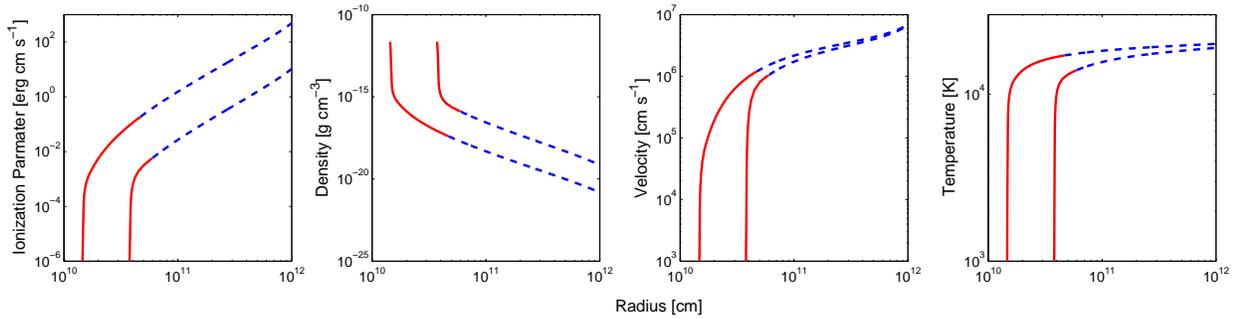}
\caption{Profiles of the flow structure for an X-ray heated flow using the realistic temperature-ionization parameter relation, for two planetary radii of $\sim 19.8$ \& $\sim 55$ $R_\oplus$ ($\approx 1.3$ \& $3.5\times10^{10}$ cm) for a Jupiter mass planet at 0.1 AU around a solar mass star with $L_X=1\times10^{30}$ erg s$^{-1}$. The solid line shows the sub-sonic part of the flow, while the dashed line shows the super-sonic portion. We note all flow structures are similar to those shown here.}\label{fig:xi1}
\end{figure*}

To determine the position of the sonic surface for a given planetary radius we can take advantage of the insensitivity of the flow solutions to the details of the underlying (non-X-ray heated) atmosphere, and instead choose a sonic radius and solve for the planetary radius as the location where the pressure in the X-ray flow diverges (as in Eq.~\ref{eqn:RpRs}).  The correct solution, and sonic surface position, is thus that which reproduces the desired planetary radius.  The accuracy of this method has been tested against more detailed (and computationally expensive) calculations that directly account for dynamical balance in the upper atmosphere of the planet.  We find that the error in the sonic radius position, and mass-loss rate, is small ($<1\%$) over large, order of magnitude changes in both temperature and planetary density, covering a range of parameter space larger than expected for exoplanets.  In addition we find that the sonic surface always occurs within optically thin, X-ray heated, gas ($N(R_s)<10^{22}$ cm$^{-2}$), validating the ionization parameter method.

Examples of such solutions are shown in Figure~\ref{fig:xi1} for two planetary radii ($\approx 1.3$ \& $3.5\times10^{10}$ cm/ 19.8 \& 55 $R_\oplus$) for a 1 Jupiter mass planet at a separation of 0.1~AU from a solar mass star with $L_X=10^{30}$~erg~s$^{-1}$, although all flow solutions show very similar features.  We again note the steep gradients close to the planet surface, and the sonic point occurs at only a few planetary radii in both cases, while at large radii the flow is approximately isothermal and shares many properties with Parker's thermal wind (\citealt{parker1958}).
\begin{figure}
\centering
\includegraphics[width=\columnwidth]{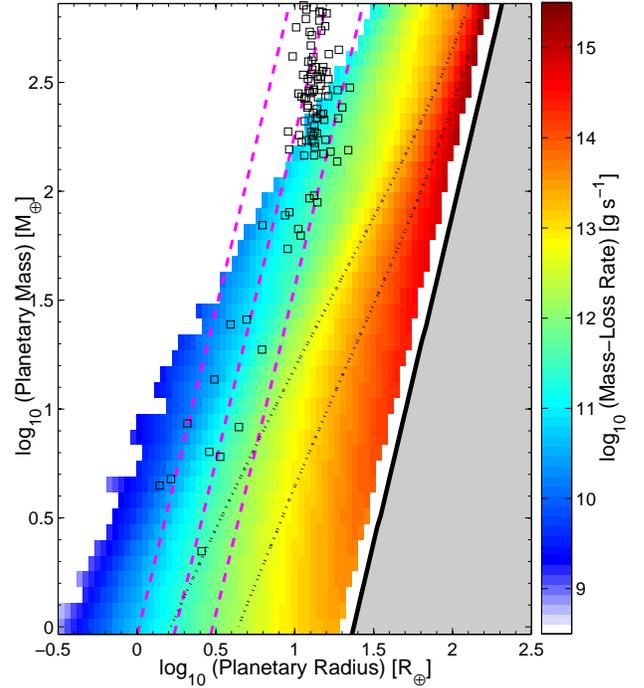}
\caption{Mass-loss rates for planets with solar metallicity, located at 0.1~AU around a solar mass star with $L_X=10^{30}$~erg~s$^{-1}$. The thick solid line shows the maximum radius at which a flow can occur (Roche radius), with the grey shaded region corresponding to dynamical Roche-lobe overflow.  The magenta dashed lines are lines of constant average density; 0.2, 1, 5~g~cm$^{-3}$ (Jupiter and Earth have densities 1.3~g~cm$^{-3}$ and 5.5~g~cm${-3}$ respectively), and the black dotted lines are `evaporation lines' for ages of 100~Myr and 1~Gyr. The black squares show a sample of currently observed transiting planets as a guide to the range of masses and radii observed for planets.}
\label{fig:30}
\end{figure}
Given that the flow solutions are specified uniquely as a function of the ionization parameter, $\xi$, along the streamline, and the flows are highly insensitive to the atmospheric structure, we can use similarity arguments to extract exact scaling relations for several of the flow variables. Since $\xi\propto L_X/\rho$ and solutions with the same $\xi(r)$ are topologically identical for fixed parameters - $M_p$, $R_p$, $a$ \& $M_*$ - then $\xi(r)$ is fixed along the streamlines if  $\rho\propto L_X$. In the case where the sonic surface occurs far from the Roche radius, which is almost always the case for realistic planetary densities (so the effect of stellar gravity and the centrifugal force is small), flow solutions with the same $\xi(r)$ profile are topologically identical for fixed parameters ($M_p$,$R_p$). Therefore, the density in the flow then scales as $\rho\propto a^{-2}$ in order to yield a fixed $\xi(r)$ profile along the streamline. Furthermore, we can include a metallicity scaling using the $\xi\propto Z^{-0.77}$ result mentioned in Section~\ref{sec:heat}. Thus we similarly find $\rho\propto Z^{-0.77}$ in a flow with fixed $\xi(r)$. Combining these we find a mass-flux that scales as
\begin{equation}
\label{eqn:xscale}
\dot{m}_X\propto \frac{L_X}{a^2Z^{0.77}},
\end{equation}
provided $R_p$ is not close to the Roche radius, however, exact scaling relations do not exist for $R_p$ and $M_p$ and these must be calculated numerically.  It is worth pointing out that our definition of `close' to the Roche-radius is somewhat different to what is meant in previous energy-limited calculations (e.g. \citealt{erkaev2007}), as the hydrodynamic solution actually depends on the gradient of the effective potential (Eq.~A2) rather than its absolute value. So by `close' we mean $R_{\rm roche}-R_s\la 0.1 R_p$, which, given $R_s$ is similar to $R_p$ and most planets do not have radii that fill their Roche-lobes, means Roche-lobe effects to the scalings of these hydrodynamic solutions will only matter in the most extreme cases.  

Importantly, as mentioned above, we must also check that our flow solutions are in the fluid limit.  Physically we require that the exospheric radius (where the fluid is no longer pressure dominated) is at a radius greater than the sonic radius, since anything that happens after the sonic point is not in causal contact with the sub-sonic portion of the flow and cannot affect it. To do this we compare the pressure scale height $\ell=\partial r/\partial \log P$ up to the sonic surface with the collisional mean free path of the gas particles given by $\lambda=1/n\sigma$. If $\ell/\lambda >1$ up to the sonic surface then the flow will be in the fluid limit, whereas if $\ell/\lambda <1$ then the calculated fluid solution is not appropriate and mass-loss is likely to occur from a ballistic type escape mechanism (e.g. \citealt{volkov2011}).

In Fig.~\ref{fig:30} we show a slice through the $M_P$ and $R_p$ plane, showing the mass-loss rates as a colour map.  The thick solid line indicates the Roche radius, the maximum radius at which a flow can occur, with any planet beyond this in the grey shaded region undergoing dynamical Roche-lobe overflow.  One consequence of $R_s$ being slightly larger than $R_p$ is that there is a range of planetary radii close to the Roche-radius where no transonic solutions exist (white-space between the colour map and thick black line), as any sonic surface would have to occur outside the Roche-radius, which as discussed above is unphysical. In reality we hypothesise that a planet in this region may drive a purely sub-sonic wind to the Roche-radius and undergo a thermally driven Roche-lobe overflow in this way. It is however important to realise that this range of parameter space is extremely small.  The white region at the left of the figure indicates the region where hydrodynamic flow solutions do not exist and instead mass-loss may occur via a ballistic type mechanism.  This figure clearly indicates that the most interesting regime for planetary evaporation is the Neptune/super-Earth regime, and that in some cases planets of Jupiter mass and above may not experience hydrodynamic evaporation at a separation of 0.1~AU.

We also show two `evaporation lines' in Fig.~\ref{fig:30} at times of 100~Myr and 1~Gyr for which we set $\dot{m}_X t=M_p$. A planet close to these lines can thus be expected to have lost a substantial fraction of its mass in 100~Myr or 1~Gyr respectively.  As a guide we show a sample of close-in planets from taken the Exoplanets Encyclopedia\footnote{www.exoplanet.eu, see also \citet{schneider2011}} on 13 September 2011 (see \citealt{jackson2012} for a more detailed description of this sample).  It is important to note though that we have not accounted for the true orbital distances of the planets here, they are intended purely as a guide for the range of masses observed (see Section~\ref{sec:evol} where we also take into account the separations).


\begin{figure*}
\centering
\includegraphics[width=0.8\textwidth]{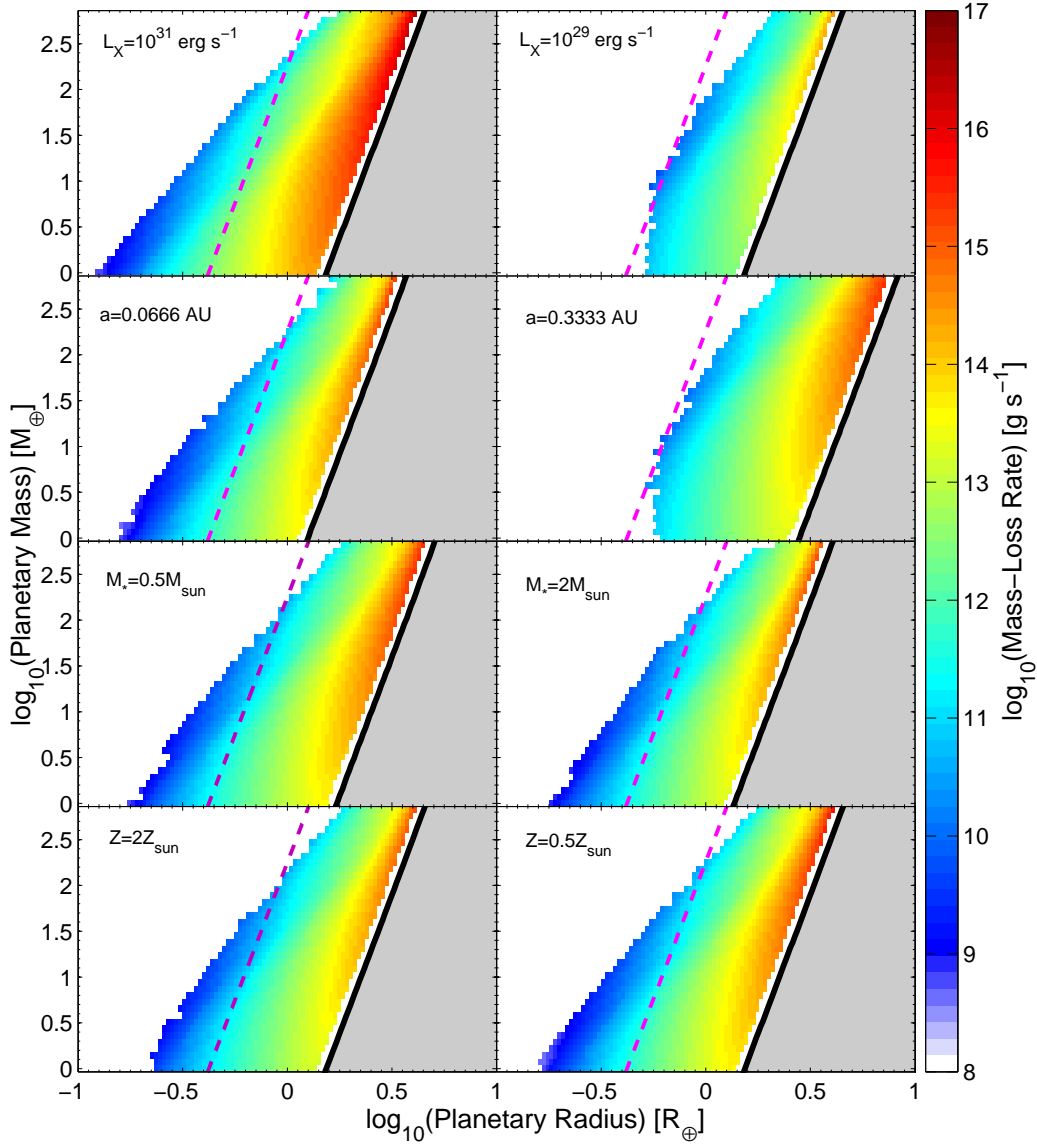}  
\caption{Mass-loss rates in the $M_P$-$R_P$ plane, varying the input parameters $L_X$, $a$, $M_*$, and $Z$.  In each panel all parameters except the one indicated is held fixed at $L_X=10^{30}$~erg~s$^{-1}$, $a=0.1$~AU, $M_*=1 M_\odot$ and/or $Z=1 Z_{\odot}$. As in Fig.~\ref{fig:30} the thick solid line indicates the Roche radius, while the dashed line indicates a mean planetary density of 1~g~cm$^{-3}$.}
\label{fig:Xgrid}
\end{figure*}

Including all the effects described above we can calculate flow solutions for a range of input parameters and we show the variation of the resultant mass-loss rates in Fig.~\ref{fig:Xgrid}, where we plot colour maps in the $R_p-M_p$ plane for various values of $a, L_X, M_*$, and $Z$.  Fig.~\ref{fig:Xgrid} shows that, as expected, planets of larger radius for the same mass have larger photo-evaporation rates, with the largest values occurring for planetary radii just inside the Roche-radius. Furthermore, at higher X-ray luminosity and smaller separations the hydrodynamic limit of evaporation extends to smaller planetary radii.  In addition, as in Fig.~\ref{fig:30}, we see that while Jupiter mass planets may undergo evaporation for a small range of parameter space, Neptune and super-Earth type planets with sensible densities can experience a reasonable level of hydrodynamic evaporation during their lifetimes.

We can compare our results here to the previous energy-limited evaporation models summarised by Eq.~\ref{eqn:elimited}, and find that Eq.~\ref{eqn:elimited} performs poorly in reproducing our results unless the energy efficiency ($\eta$) and expansion radius ($\beta$) are allowed to vary by large amounts (in some cases orders of magnitude).  Additionally, as we discuss in Section~\ref{sec:discuss} we find that X-ray evaporation is not actually occurring in the energy-limited regime.  Thus, while the mass-loss in the X-ray driven regime does have the same scaling with X-ray luminosity and separation (Eq.~\ref{eqn:xscale}), we find that Eq~\ref{eqn:elimited} is of limited use for X-ray driven evaporation, particularly when considering wide ranges in planetary mass and radius.

\subsection{EUV driven}
\label{sec:euv}
In the case of EUV driven evaporation, the ionization front is below the X-ray sonic surface and thus any X-ray flow is sub-sonic. As discussed previously EUV driven evaporation can exist in two limits where ionization is balanced by either gas advection (`energy-limited') or recombination (`recombination-limited').  At high fluxes, similar to the values during the star's early evolution, EUV evaporation can be considered to be recombination-limited (\citealt{murrayclay2009}). In order to assess the impact of evaporation on the planet mass, at this early stage we are concerned with the high luminosity phase of the stellar evolution and consider EUV driven evaporation to be in the recombination-limited regime.  In this case the ionization front is set by balancing ionization against recombination.

An ionization front can be of two types; when the speed on the neutral side $v_{\rm in}>2c_{\rm EUV}$ the ionization front is a rarefied (R-type) front, whereas when $v_{\rm in}<c_X/2(c_X/c_{\rm EUV})$ the ionization front is a detached (D-type) front (Spitzer 1978). In most cases $c_{\rm EUV}\gg c_X$, so the ionization front will be D-type and there will be a density jump across the ionization front with a maximum input speed $v_{\rm in, max}=c_X/2(c_X/c_{\rm EUV})$. As such the neutral X-ray flow will be highly sub-sonic and the density structure will be approximately described by the hydrostatic structure. 

As described in Section~\ref{sec:X1}, such a hydrostatic structure will be very steep near the planet surface.  As such we cannot assume that the sub-sonic flow occurs over of order one scale height or use a total column argument to determine its width as in \citealt{johnstone1998}. Furthermore, the speed on the ionized side of the ionization front will depend on escape temperature at the ionization front (see Figure~\ref{sfig:EUV}). If the escape temperature is less than $10^4$K at the ionization front, then there is no sonic point in the flow and the speed on the ionized side of the front is $v_{\rm out}\sim c_{\rm EUV}$ (\citealt{johnstone1998}). However, should the escape temperature at the ionization front be $>10^4$~K then the speed on the ionized side of the front will be described by the Parker wind solution ($v_{\rm out}=f_{\rm parker}c_{\rm EUV}$), as a result of the necessity that an isothermal flow has a sonic point at $R_s^{-1}=1/(2c_{\rm EUV})\left.\partial \Psi_{\rm eff} / \partial r \right|_{r=R_s}$ (\citealt{parker1958,parker1960}).

In the recombination limit we are considering here the position of the ionization front is set by balancing ionization and recombination along a line from the star to the planet, so
\begin{equation}
\label{eqn:ion1}
\frac{\Phi_*}{4\pi a^2}=\int_a^{R_{\rm IF}}\alpha_r n_{\rm EUV}^2=A \alpha_r n_{\rm EUV}^2R_{\rm IF}.
\end{equation}
Now writing $R_{\rm IF}=(1+x)R_p$, where $x$ encapsulates the stand-off distance of the ionization front due to the partially neutral X-ray flow behind it, the total mass-loss rate may be written as $\dot{m}_{\rm EUV}=4\pi\mu n_{\rm EUV}f_{\rm parker}c_{\rm EUV}R_{\rm IF}^2$.  Evaluating this at the ionization front we obtain the following equation for the total mass-loss rate in EUV driven planetary photoevaporation,
\begin{equation}
\label{eqn:euvmdot}
\begin{split}
\dot{m}_{\rm EUV}&=2.4\times10^{11}\textrm{g s}^{-1}(1+x)^{\frac{3}{2}} f_{\rm{parker}}
                      \left(\frac{\Phi_*}{10^{40}{\rm s}^{-1}}\right)^{\frac{1}{2}}
                      \\
& \times \left(\frac{a}{0.1{\rm AU}}\right)^{-1}
                \left(\frac{R_p}{10 R_{\oplus}}\right)^{\frac{3}{2}}
                \left(\frac{A}{1/3}\right)^{\frac{1}{2}}
                \left(\frac{c_{\rm EUV}}{10{\textrm{km s}^{-1}}}\right).
\end{split}
\end{equation} 
We note that Eq.~\ref{eqn:euvmdot}, appears throughout the literature for EUV driven transonic winds.  For the case of $f_{\rm parker}=1$ and $x=0$ this solution is identical to the well known solution for the evaporation of gas clouds of radius $R_p$ presented by \citet{bertoldi1990}.  This solution also appears for externally irradiated protoplanetary disks where $f_{\rm parker}=1$ (\citealt{johnstone1998}), and for hot Jupiters irradiated with a high luminosity pure EUV spectrum with $x=0$ (\citealt{murrayclay2009}).  As can be clearly seen, in the case of EUV driven evaporation there is a qualitatively different scaling with the stellar high-energy luminosity and the orbital distance than for X-ray driven evaporation ($\dot{m}_X\propto L_X/a^2$). This indicates that at large separations and low luminosities it will be the EUV rather than the X-ray that is dominating the evaporation rates.
\begin{figure*}
\centering
\includegraphics[width=\textwidth]{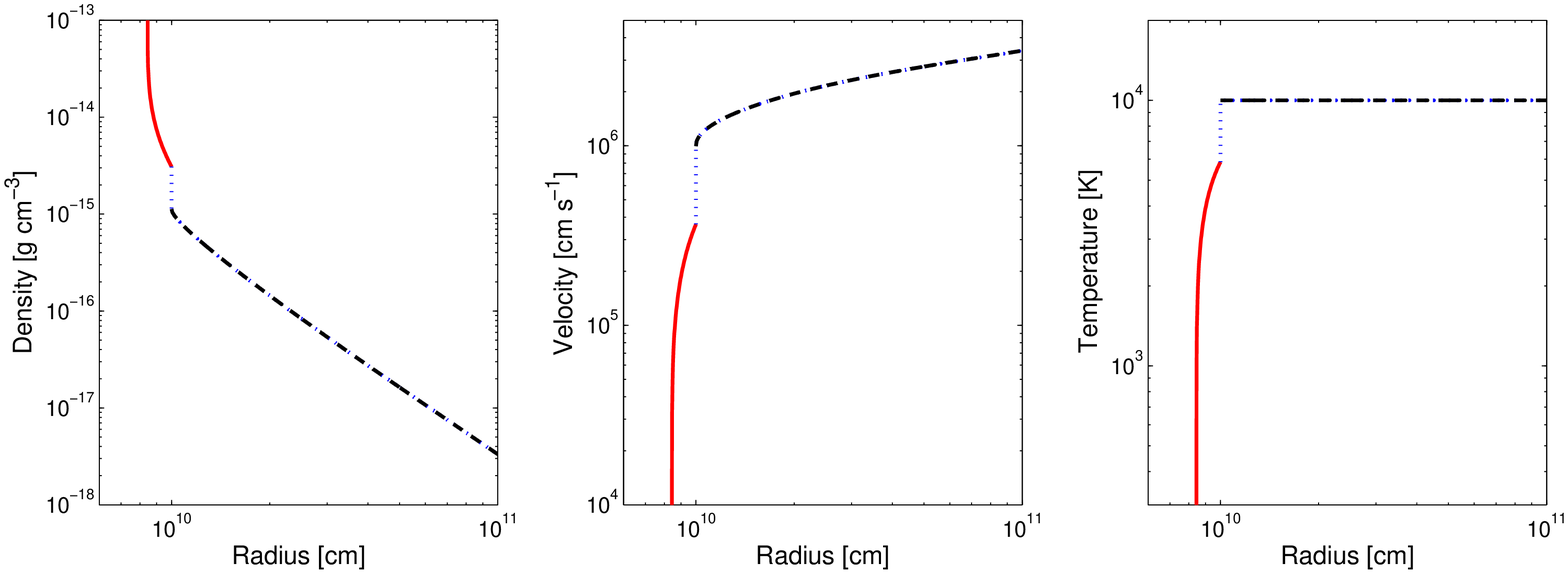}
\caption{The structure of an EUV driven flow, with a sub-sonic X-ray heated flow at smaller radius. The X-ray heated flow is shown by the solid line, the EUV heated flow by the dashed line and the ionization front by the dotted line. The panels show density, velocity and temperature (from left to right), as a function of radius. The planet is 1~$M_J$ at a separation of 0.1~AU, with a radius of 12.9~$R_\oplus$ ($8.2\times10^9$~cm), an X-ray luminosity of $L_X=10^{29}$~erg~s$^{-1}$ and an EUV luminosity of $\Phi_*=10^{40}$~s$^{-1}$.}
\label{fig:EUV1}
\end{figure*}
To find $x$ we must solve the jump conditions across the ionization front. Using mass and momentum conservation, and writing the pressure as $P_X=c_X^2\rho_X$ on the X-ray heated side and $P_{EUV}=c_{\rm EUV}^2\rho_{\rm EUV}$ on the EUV heated side, the deterministic equation for the density behind the ionization front becomes
\begin{equation}
\begin{split}
f_{\rm parker}T_{\rm EUV}\left(\frac{n_{EUV}}{n_X}\right)^2 &+ f(L_X/n_Xa^2) \\
& \hspace{-10ex} - T_{\rm EUV}\left(1+f_{\rm parker}^2\right)\left(\frac{n_{EUV}}{n_X}\right)=0,
\end{split}
\end{equation}
where $n_{\rm EUV}$ is given by Eq.~\ref{eqn:ion1}.  This equation possesses two roots, although only one can correspond to an out-flowing, sub-sonic X-ray heated flow\footnote{There always exists the possibility of an inflowing solution.}.  Once $n_X$ is found a Bernoulli potential can be constructed for the sub-sonic portion of the flow, and $x$ can then be found in an identical fashion to that used to find the full X-ray solutions in Section~\ref{sec:X1}.  In Figure~\ref{fig:EUV1}, we show an example solution of an EUV driven flow above a sub-sonic X-ray flow from a Jupiter mass planet.

\begin{figure*}
\centering
\includegraphics[width=0.8\textwidth]{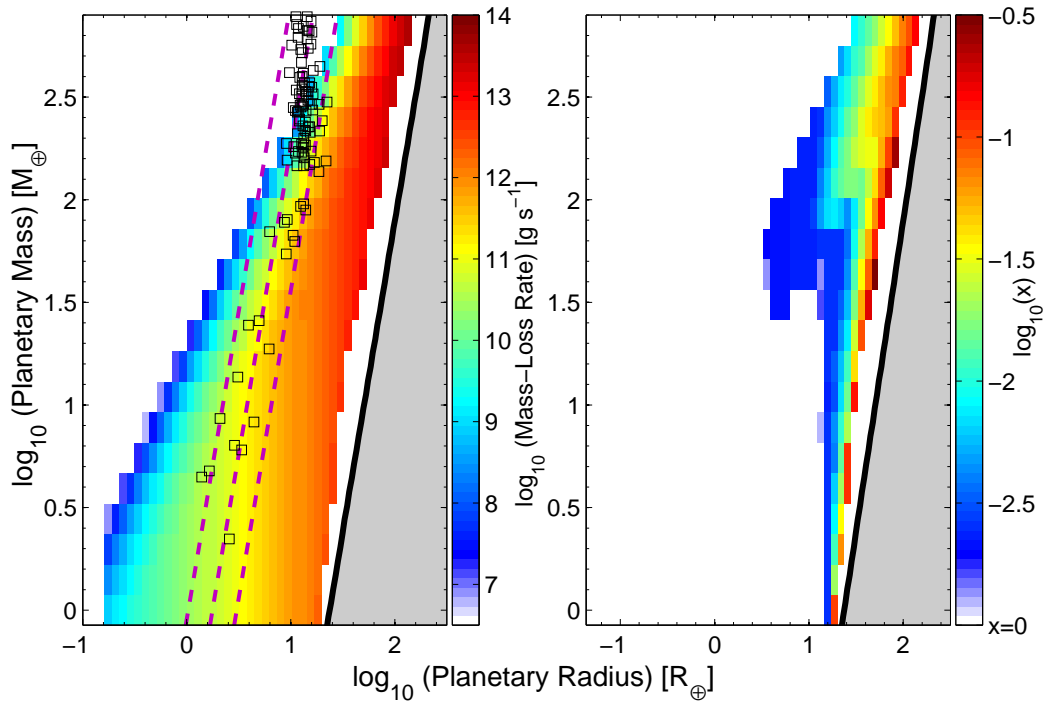}
\caption{Mass-loss rates (left) and stand-off distance of the ionization front (right) in the $M_P-R_P$ plane. As in Fig.~\ref{fig:30} the planet is solar metallicity at 0.1~AU from a solar mass star and in the left hand panel we include the same lines of constant density and sample of known transiting exoplanets.  The star has an EUV luminosity, $\Phi_*=10^{40}$~s$^{-1}$, while the X-ray luminosity is lowered compared to Fig.~\ref{fig:30} to $L_X=10^{29}$~erg~s$^{-1}$ to increase the parameter space over which the EUV dominates the evaporation. }
\label{fig:euv2}
\end{figure*}

In order to consider the role the X-ray radiation plays in enhancing the mass-loss rate we show both the total mass-loss rates and the value of $x$ for a range of parameter space in Fig.~\ref{fig:euv2}. This figure shows the mass-loss rates and stand-off distances from planets located at 0.1 AU around a solar mass star with $\Phi_*=10^{40}$ s$^{-1}$, where $L_X$ has been lowered from the nominal value used earlier in this work to $10^{29}$ erg s$^{-1}$, to increase the parameter space over which the EUV dominates the evaporation. We note that in most cases $x\ll 1$ although $x$ does become of order unity at higher values of $L_X$ and larger planetary radii. In this case the white region between the colour-map and the Roche radius is in part due to EUV driven solutions not being possible in this region as the flow has transitioned from EUV to X-ray dominated.  

\begin{figure*}
\centering
\includegraphics[width=0.8\textwidth]{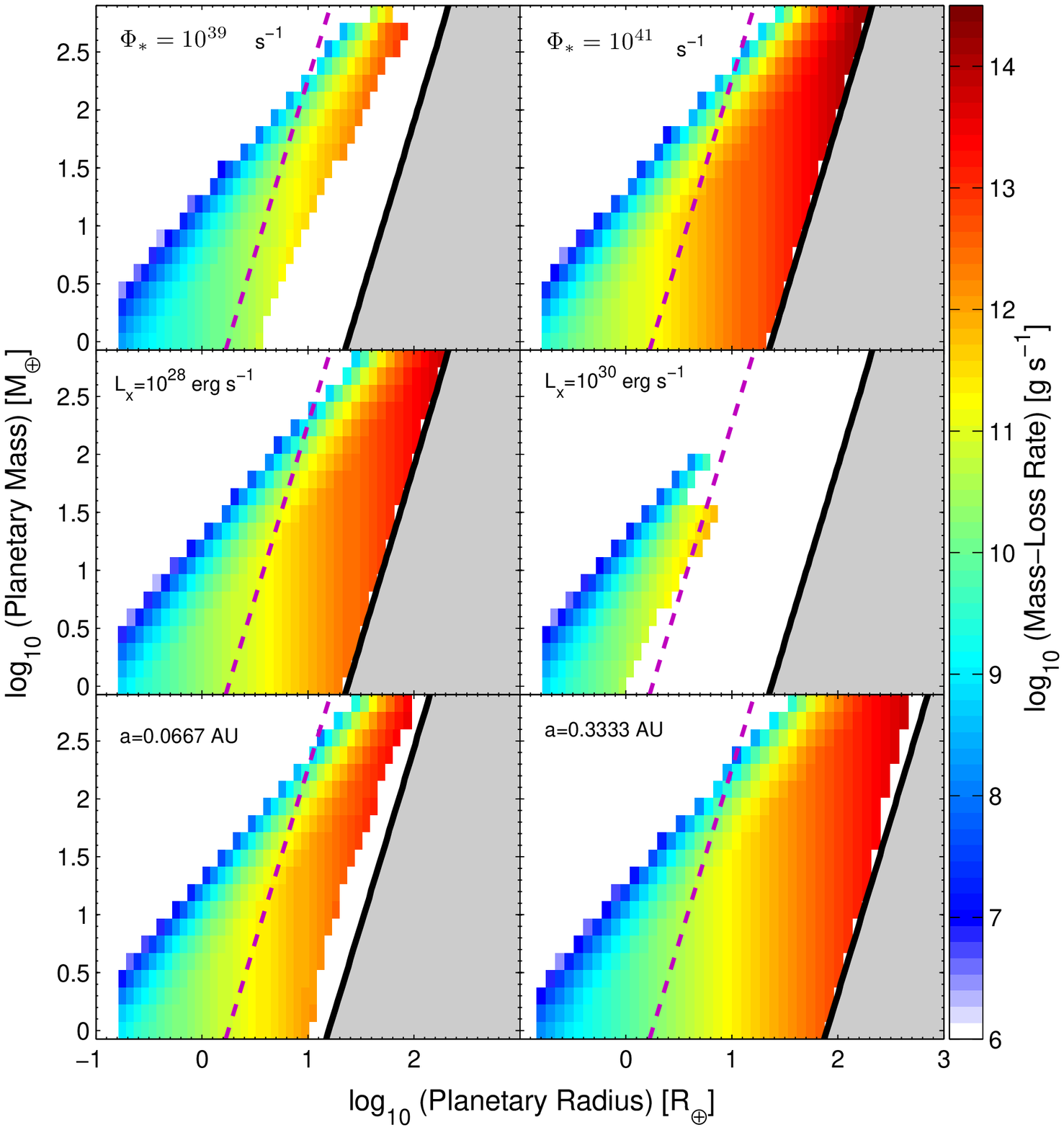}
\caption{Variation of the EUV driven evaporation rate with various input parameters. Deviations fron the standard parameters of $\Phi_*=10^{40}$~s$^{-1}$, $L_X=10^{29}$~erg~s$^{-1}$, $a=0.1$~AU and $M_*=1M_{\odot}$ are shown in the panels.  Lines as described in Figure~\ref{fig:Xgrid}.  White space between the colour-map and the Roche lobe is largely due to transition from EUV domination to X-ray domination.}
\label{fig:euvbig}
\end{figure*}

To explore the parameter space further Fig.~\ref{fig:euvbig} shows the variation of the mass-loss rates and region of EUV dominance with $\Phi_*$, $L_X$, and $a$, where deviations from the input parameters used in Fig.~\ref{fig:euv2} are shown in the panels. This shows that, as expected due to the different scalings of the mass-loss rate from EUV ($\dot{m}_{\rm EUV}\propto \Phi_*^{1/2}/a$) and X-ray ($\dot{m}_X\propto L_X/a^2$) dominated flows, EUV photo-evaporation dominates at low X-ray luminosities and large separations, whereas X-ray photo-evaporation dominates at high X-ray luminosities and small separations.  As a result X-ray dominated winds will in general lead to larger mass-loss rates than EUV dominated winds.
    
\subsection{Direct comparison to HD209458b}
{HD209458b has an extended atmosphere which has been well studied in the past, particularly using pure EUV evaporation.  \citet{penz2008} estimated an X-ray luminosity of $\sim 1.1\times10^{27}$ erg s$^{-1}$ and an EUV luminosity of $\sim 6\times10^{37}$ s$^{-1}$ (their $L_{\rm XUV}-L_X$) for HD209458b, such a combination places it firmly in the EUV driven regime as an X-ray luminosity of $\sim10^{29}$ erg $^{-1}$ would be required to switch to the X-ray driven regime. Therefore, we indicate that the previous pure EUV approach for modelling HD209458b's evaporation is probably accurate and our models indicate a mass-loss rate of $\sim 2\times10^{10}$ g s$^{-1}$ in good agreement with previous EUV only models and observations of HD209458b \citep{vidalmadjar2003,vidalmadjar2004}.  Additionally we find the X-rays only make a small difference to the flow structure at small radius with an sub-sonic X-ray heated flow of size $\la 0.01 R_p$. Given the very small correction the inclusion of X-ray heating makes when considering the current evaporation of HD209458b in this case our conclusions essentially reduce to a pure EUV flow similar to those calculated by \citet{murrayclay2009}.  As a result of the simplifying assumptions made in the treatment of the EUV in this work, previous EUV only models that include more detailed chemistry (e.g. \citealt{koskinen2010a, benjaffel2010}) may better describe the details of the flow solution in this case. At early times however, HD209458b will have undergone evaporation in the X-ray dominated regime with mass-loss rates $\sim 10^{13}$ g s$^{-1}$, several times larger than the predictions of previous EUV models, indicating that HD209458b may have lost up to $2-5\%$ of it's original mass, again somewhat larger than previous EUV only models (\citealt{murrayclay2009} found $\sim0.6\%$), but certainly not enough to have made it unstable to complete destruction.}

{In the case where detailed models show that the heating is EUV dominated and predominately takes place in a small layer near the base of the flow, Eqn.~\ref{eqn:elimited} may be used to provide reasonable estimates of the mass-loss rate with an appropriately chosen efficiency (as in the case of HD209458b discussed above $\eta \sim 25\%$ e.g. Yelle et al. 2004).  However, we caution that the heating efficiency certainly does not remain constant for the parameter space considered in this work and as such care must be taken when using Eqn. \ref{eqn:elimited} to estimate the mass-loss rate. Where a choice of an appropriate `efficiency' requires detailed calculations, such as in this work or others (e.g. Yelle et al. 2004; Murray-Clay et al. 2009).}

\section{Evolution}
\label{sec:evol}
\begin{figure}
\centering
\includegraphics[width=\columnwidth]{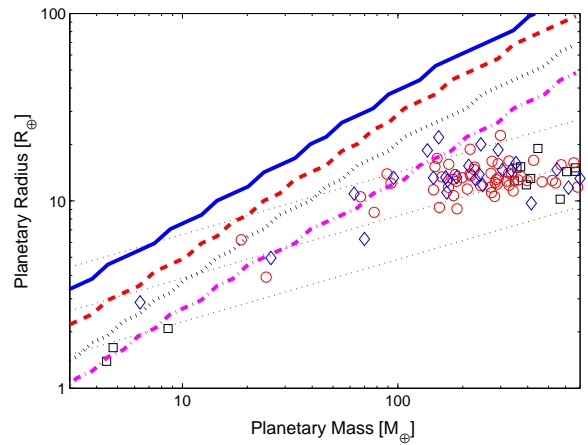}
\caption{Crude planetary survival lines at different separations compared to the observations. The points and lines are coded correspondingly: the solid blue line indicates the planetary survival line at 0.1~AU and blue diamonds the known planets with separations 0.05-0.1~AU, the survival line for 0.05 AU is red dashed and planets at 0.025-0.05~AU are red circles, the survival line at 0.025AU is black dotted and planets at 0.0125-0.025 AU are black squares, and the survival line for 0.0125AU is magenta dot-dashed while there are no planets at $<0.0125$~AU. Planets above their respective survival line in the figure should have been destroyed according to the models.  We also show lines of constant density as a guide at densities of $0.2$, 1 and 5~g~cm$^{-3}$.}
\label{fig:surv}
\end{figure}

To assess the effect of evaporation on the planet's evolution in this section we build simple evolutionary models to indicate whether there are any regions of parameter space in which evaporation can significantly influence a planet over its lifetime.  To do this we follow the evolution of the planet mass through time, as well as the evolution of the stellar high-energy spectrum.  We follow \citet{jackson2012} to include the evolution of the stellar high-energy spectrum, using their results for the evolution of the X-ray component of the spectrum.  After an initial phase ($t<\tau_{\rm sat}$) where $L_X$ is saturated at a level $L^{\rm sat}_X$, the X-ray luminosity then decays as
\begin{equation}
L_X(t)=L^{\rm sat}_X\left(t/\tau_{\rm sat}\right)^{-\alpha}
\end{equation}
where values of $L^{\rm sat}_X$, $\tau_{\rm sat}$ and the power law index $\alpha$ for different spectral types are listed in Table~3 of \citet{jackson2012}. In the case of pure X-ray evaporation we can use the results of Section~\ref{sec:X1} to estimate a planetary survival line at $\sim$3~Gyr for different separations under the very crude assumption of evolution at constant mass and radius (see below for models that follow the evolution of the planet mass), to compare to observations. This is shown in Fig.~\ref{fig:surv}, where models at four separations, 0.0125, 0.025, 0.5 and 0.1 AU, are shown. Any planet above the corresponding survival line should have completely lost is envelope to evaporation at an age of $\sim$3~Gyr, while a planet on or just below the survival line would indicate that it has experienced a period of strong evaporation.  Planets well below the survival line would indicate that while the planet might be evaporating, it has not changed significantly from its initial mass.

The observations indicate that, as expected if our models are accurate, no planets lie above their survival line.  As mentioned above the observations also indicate that the most interesting regime for planetary evaporation is the hot-Neptune regime, and that the hot-Jupiters, while evaporating, do not loose substantial fractions of their initial mass.  

\subsection{Transition from X-ray to EUV driven}
\label{sec:transition}
The evolution of the EUV luminosity is still observationally rather unclear.  Motivated by the fact that EUV luminosities and X-ray luminosities seem to have similar values at earlier times during the T-Tauri phase (\citealt{alexander2004}) and at late times (\citealt{ribas2005}), perhaps - indicates fall off with time is similar, if not identical, (see also \citealt{sanzforcada2011}), we make a simple assumption and couple the X-ray and EUV luminosity to have identical values at all times. Thus we take the EUV luminosity to fall in step with the X-ray luminosity.

\begin{figure}
\centering
\includegraphics[width=\columnwidth]{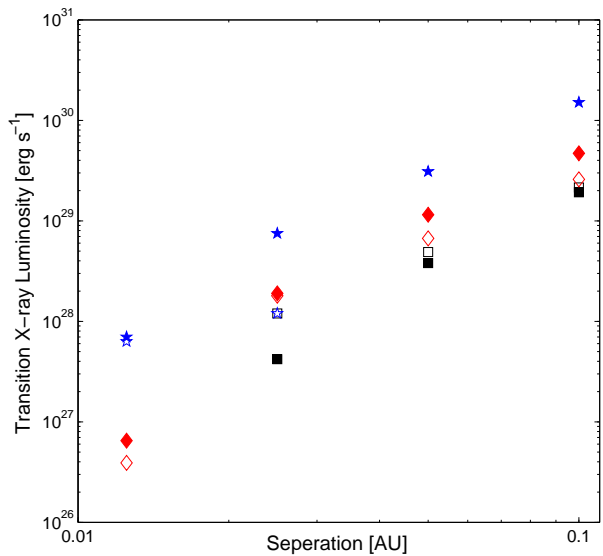}
\caption{The X-ray luminosity at which a planetary wind will transition from being X-ray driven to EUV driven is shown for different separations. The points indicate Jupiter mass planets (unfilled) and Neptune mass planets (filled), with densities of 0.1 (black squares), 0.3 (red diamonds) \& 1 (blue stars) g~cm$^{-3}$.}
\label{fig:tranx}
\end{figure}

The transition from X-ray driven to EUV driven evaporation is both of observational and theoretical interest. Using the models calculated in Section~\ref{sec:basichydro} we can estimate the X-ray luminosity at which a planet of given mass, radius and separation will switch from becoming X-ray driven to EUV driven. The results of these calculations are shown in Fig.~\ref{fig:tranx}. We note that the X-ray luminosity at which this transition takes place for Jupiter mass planets with densities of $\sim$1~g~cm$^{-3}$ at separations of $\ga 0.05$ AU are considerably higher than any expected X-ray luminosity for a solar-type star and are not shown.   As expected, at higher planetary densities, and larger separations, the transition takes place at higher X-ray luminosities.  In Section~\ref{sec:Ill} we will see that this has implications for whether the EUV or X-ray will drive the evaporation at late ($\sim$Gyr) times, at which we observe most planets.

\subsection{Illustrative Cases}
\label{sec:Ill}
The above survival calculations are presented as a guide to indicate which regimes are likely to be the most interesting, in terms of evaporation having an important role on planetary evaporation. In this section however we construct more accurate models that account for the evolution of the planet mass as well as the transition from X-ray driven to EUV driven evaporation. Therefore, we consider the two cases: that of a Jupiter mass planet and that of a Neptune mass planet, at orbital separations of 0.0125, 0.025, 0.05 and 0.1~AU, and planetary densities of 0.1, 0.5 and 1~g~cm$^{-3}$.  Following the radial evolution of the planet is beyond the scope of this work, since this would require coupled modelling of the planetary interior, and thus we take the planets to evolve at constant density.

Furthermore, should the planet transition from X-ray driven to EUV driven, we also check to see whether the EUV evaporation is in the `recombination limit' or the `energy limit' using the results of \citet{murrayclay2009}.  Specifically, if the recombination-limited EUV flow gives a mass-loss rate above that of the energy-limited case, we switch to the energy-limited formula for EUV evaporation provided by \citet{murrayclay2009}.  In the energy-limited case the mass-loss rate will now scale linearly with the ionizing luminosity, rather than as the square root as in the recombination-limited case.

\begin{figure*}
\centering
\includegraphics[width=0.9\textwidth]{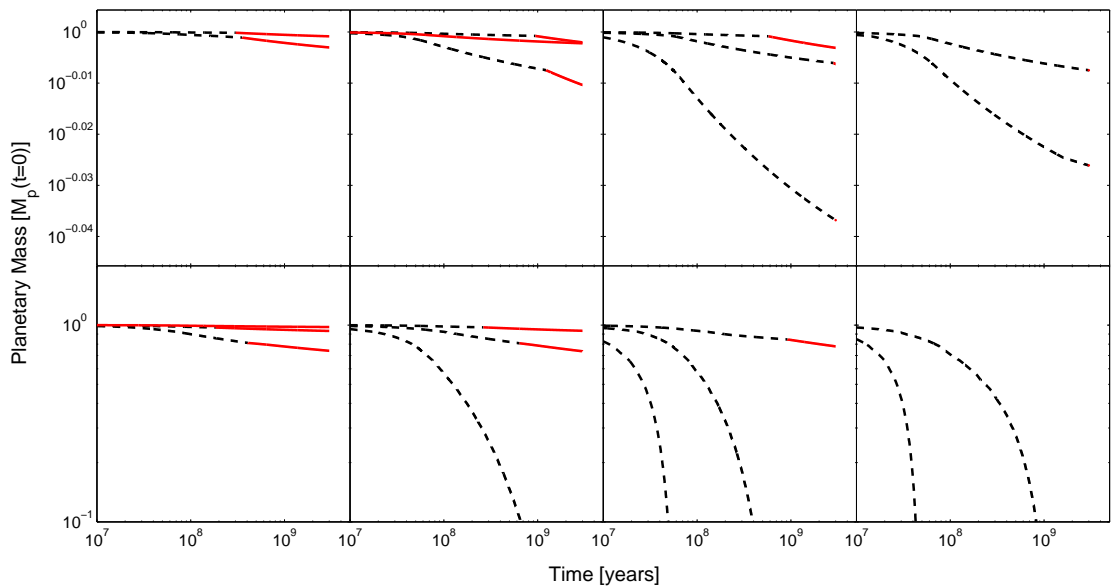}
\caption{The mass evolution of close-in planets for initially Jupiter mass planets (upper panels) and Neptune mass planets (lower planets) at densities of 0.1, 0.3 \& 1 g cm$^{-3}$.  Planets of lower density lose mass faster.  Note that planets of density 0.1~g~cm$^{-3}$ at 0.0125~AU and Jupiter mass planets of density 1~g~cm$^{-3}$ at 0.1~AU are not shown as the flow is not in the hydrodynamic limit.  From left to right the panels indicate decreasing separations from 0.1, 0.05, 0.025 to 0.0125~AU. Dashed black lines indicate the time when the evaporation is dominated by X-rays and solid red lines when the evaporation is dominated by the EUV.}
\label{fig:evolveind}
\end{figure*}

The results of these evolution calculations are shown in Fig.~\ref{fig:evolveind}, where the upper panels show planets with an initial mass of one Jupiter mass, and the lower panels show planets with an initial mass of one Neptune mass with separation decreasing from left to right.  Note that planets with a density of $0.1$~g~cm$^{-3}$ at a separation of 0.0125~AU are not shown as these planets would be undergoing dynamical Roche-lobe overflow rather than evaporation. In addition, a Jupiter mass planet with a density of 1~g~cm$^{-3}$ at 0.1~AU does not experience hydrodynamic evaporation, and is not shown. 

These calculations show that Jupiter mass planets experience relatively little evaporation, losing no more than 10 per cent of their initial mass, even at the closest separations.  However, Neptune mass planets are prone to destruction at separations $\la 0.05$~AU and can loose several tens of per cent of their initial mass at separations of $0.05-0.1$ AU. Thus we would not expect to see Neptune mass planets with low densities at small separations, as indicated by the observations (e.g. \citealt{hansen2011}).  At late times we also notice a clear distinction in the driving mechanism as a function of separation. At separations $\la 0.025$~AU the evaporation of Jupiter mass planets appears to be X-ray driven while at larger separations it is EUV driven.  This prediction could be compared to observations of the neutral content of the wind as a function of separation, though it must still be remembered that an X-ray driven flow can still have an EUV heated and fully ionized region above the sonic surface.       
 
\subsubsection{Comparison with previous studies}
{There have been several previous attempts to quantify the role of evaporation on the evolution of close in planets, Lammer et al. (2009) use the `energy-limited' formalism for pure EUV evaporation, \citet{penz2008} use fixed heating efficiency models that include X-ray and EUV luminosity and \citet{jackson2012} use the `energy-limited' formalism to assess the role of pure X-ray evaporation.}

{Lammer et al. (2009) estimated that pure EUV evaporation with an efficiency of $10\%$ was not capable of completely destroying most close in planets, with Jupiter mass planets being more stable than lower mass planets, similar to what is found in this work. While the inclusion of X-ray mass-loss at earlier times does increase the total mass lost by close-in planets, it is not sufficient to make Jupiter mass planets unstable to complete evaporation.  However, we do find lower mass ($\sim$ Neptune mass) planets are more unstable to evaporation than predicted by Lammer et al. (2009). }

{\citet{penz2008} considered the evolution of hot Jupiter's similar to HD209458b and found total mass-loss over Gyr's of evolution higher than found here, this is similar to the results of certain cases in the \citet{jackson2012} models for pure X-ray evaporation. This arises for two reasons: i) the fixed heating efficiencies (60\% for \citealt{penz2008} and 25\% for \citealt{jackson2012}) are rather large when compared to the actual values obtained in certain regimes (see Section~\ref{sec:efficiency}); ii) the X-ray and EUV field were considered to both be contributing to the heating at all times, whereas we have shown there is a switch from X-ray driven to EUV driven at late times which results in reduced mass-loss rates.  Thus we conclude that estimates of total mass-loss at levels around $30\%$ for hot Jupiter's at separations of 0.01-0.02AU are unrealistically high. } 

{The use of X-ray evaporation at earlier times which switches to EUV evaporation at late times thus represents an intermediate between pure EUV evolution models, and models which include a contribution from the X-ray at all times.} 
 
\subsubsection{Limitations}
\label{sec:limitations}
The major limitation of these evolution calculations is that we do not follow the evolution of the interior structure of the planet and hence its radius.  We make the assumption the planet evolves at constant density, motivated by the fact that the spread in the density of observed planets is comparatively small relative to the orders of magnitude changes in mass.  It is, however, worth noting that a planet whose density decreases with time (as is the case for many models of planets undergoing mass loss, e.g. \citealt{baraffe2004}) will be: i) prone to higher mass-loss rates for a given mass; and ii) experience a transition from X-ray driven to EUV driven at later times. These effects will result in more rapid planetary evaporation than our calculations here.  On the other hand if the density of the planet increases with time, as may happen if a large fraction of the planet mass is comprised of a rocky core, the converse will be true and the planet may be able stabilize itself against evaporation.

In addition, \citet{baraffe2004} point out that a critical point may occur when a planet's evaporation timescale becomes shorter than the thermal timescale. In this case the planet is no longer able to shrink its radius to adjust to the mass-loss, and the planets evaporation is accelerated (lower mass with same radius results in higher mass-loss) and can lead to runaway mass-loss. This effect, particularly combined with a general tendency for density to decrease with decreasing mass, may make planets more prone to destruction compared to our simple calculations.
\section{Discussion}
\label{sec:discuss}
In the previous sections we have performed basic hydrodynamic calculations to investigate the evaporation of close-in planets, and the various limits it may exist in.  Qualitatively we have shown that most close-in planets ($a \la 0.1$~AU) will experience a period of hydrodynamic mass-loss driven by X-rays at early times, followed by EUV driven at late times, as the high-energy luminosity drops off with time. Furthermore, due to the steep nature of the X-ray heated sub-sonic flow, the atmospheric structure is unimportant in determining the flow structure, although as a consequence the flow structure will be sensitive to the planetary radius. By considering the combination of EUV and X-ray heating, we have shown that X-ray evaporation dominates when the sonic surface occurs in the X-ray heated flow, typically for high mass-planets with large radii and a high X-ray flux.  When the EUV evaporation dominates on the other hand, the X-ray portion of the flow is entirely sub-sonic, and EUV domination typically occurs for low mass-planets with small radii and low X-ray flux.

\subsection{Hydrodynamic or Ballistic Escape?}
\label{sec:hydro/ballistic}
One of the questions we set out to answer in this work is whether, in general, the evaporation of close-in planets occurs through a hydrodynamic process, or through a ballistic process (e.g. Jean's escape). It is clear from this work that the dominant mass-loss process for close-in planets ($a\la0.1$~AU) is through a hydrodynamic wind.  However above a certain mass dense planets are too massive to loose gas hydrodynamically, either through X-ray or EUV evaporation. Fig.~\ref{fig:Xgrid} and \ref{fig:euvbig} indicate that planets more massive than 1 Jupiter mass at densities greater than 1~g~cm$^{-3}$ have atmospheres which are too tightly bound to drive a hydrodynamic wind at separations $\sim$1~AU.  At very close orbital separations however, $<0.05$ AU, even planets of a few Jupiter masses are able to undergo hydrodynamic evaporation.  As the X-ray luminosity drops, so does the region which is prone to hydrodynamic evaporation, as clearly indicated by Figure~\ref{fig:Xgrid}. Given the much higher collision cross-section of gas particles in fully ionized (and therefore EUV heated) gas, planets switch to EUV driven evaporation and remain EUV driven before they become unable to evaporate hydrodynamically for all Jupiter and Neptune mass planets considered in Section~\ref{sec:evol}.

\subsection{Limitations of the models}
\label{sec:assump}
These models have given valuable insight into which processes dominate the evaporation in different regions of parameter space. However, it is important to highlight the limitations of the models and in particular discuss the two major assumptions of this work: adopting a solar composition for the X-ray heating (although we do account for an overall metallicity scaling) and assuming radiative equilibrium.

\subsubsection{Role of atmospheric composition}
\label{sec:atmoscomp}
Perhaps one of the major uncertainties when discussing X-ray flows is the role of composition, since X-ray heating and cooling is done by metals.  Although calculations of atmospheric composition are beginning to be performed \citep{yelle2004,garcia2007,koskinen2007,koskinen2010b}, as a starting point we have assumed the metals have abundance ratios comparable to solar values, and included a basic metallicity scaling derived from previous work on evaporating flows (\citealt{ercolano2010}). Thus our models are unlikely to be appropriate for considering very exotic, metal rich atmospheres or where the compositions are significantly different from solar.  The discrepancy may not be as bad as one would originally expect though, since as discussed in Section~\ref{sec:heat}, X-ray photo-electron heating, and line cooling, are typically done by the same metal species, with Oxygen and Carbon typically being the dominant species (\citealt{ercolano2008}). As Oxygen and Carbon has been detected in evaporating atmospheres of close-in planets (e.g. \citealt{vidalmadjar2004,benjaffel2010}), this is reassures us that the models presented in this work represent a good starting point. Nonetheless future work is certainly needed to determine the role atmospheric composition has on evaporation, along the lines of coupling the photo-chemistry of important species e.g. Hydrogen, Helium, Carbon and Oxygen to the hydrodynamic models. 

\subsubsection{Energy efficiency}
\label{sec:efficiency}
Estimates of hydrodynamic evaporation have previously been calculated through so called `energy-limited' evaporation where most of the received energy, assumed to be deposited at the planet surface, can be taken to be lost in hydrodynamic expansion as $P{\rm d}V$ work (e.g. Watson et al. 1981). However, there is no {\it a priori} reason to expect this, and in-fact \citet{murrayclay2009} determined a transition between energy-limited EUV evaporation at low EUV fluxes, and the case where the received energy is predominately re-emitted as cooling radiation at high EUV fluxes. Furthermore, comparisons of the heating and cooling timescales with the dynamical timescales of X-ray heated gas in this ionization parameter range indicate that the X-ray gas is close to radiative equilibrium (Owen et al. 2010).  Thus similarly for X-ray winds, the energy emitted as cooling radiation is considerably larger than that which goes into $P{\rm d}V$ work, suggesting much lower efficiencies than typically assumed using the energy-limited method.  This may not ultimately result in much lower evaporation rates however, as the X-ray wind is free to absorb energy out to its sonic surface.

\begin{figure}
\centering
\includegraphics[width=\columnwidth]{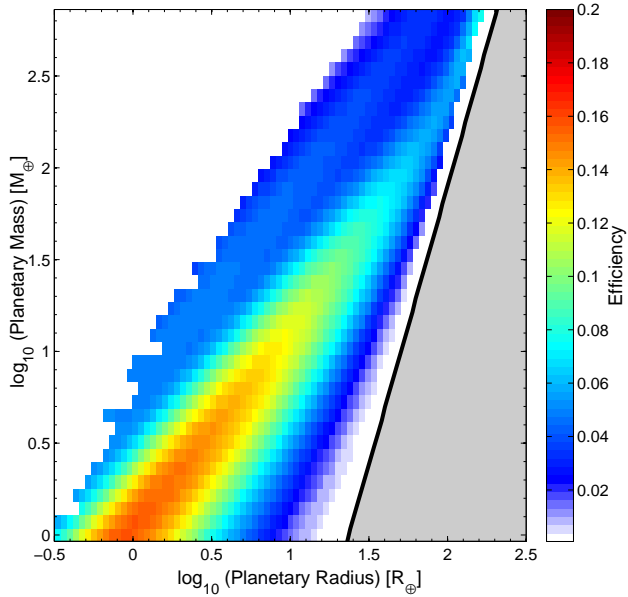}
\caption{The `efficiency' of the X-ray wind calculated by comparing the energy input received by the wind out to the sonic surface, to the mechanical luminosity of the wind [$4L_{\rm mech} a^2/(L_X R_s^2)$].  Here the calculation is performed for the X-ray flow of Fig.~\ref{fig:30}, though note that as $L_{\rm mech}\propto L_X/a^2$ this efficiency is invariant under changes to $L_X$ and $a$ provided the change in $R_s$ is small. }
\label{fig:Xeff}
\end{figure} 

In this work we have assumed that the X-ray gas is in radiative equilibrium and neglected $P{\rm d} V$ cooling based on comparisons of the dynamical and thermal timescales. However, it is prudent to check this assumption by comparing the mechanical luminosity ($L_{\rm mech}=\dot{m}_w \times \Delta e$, where $e$ is the {\it total} specific energy of the gas) of our X-ray evaporative winds to the energy input rate and calculate the `efficiency' of our flows.  An efficiency value $> 1$ would indicate that our assumption of radiative equilibrium is poor and that cooling due to $P{\rm d} V$ work is important.  In Fig.~\ref{fig:Xeff} we plot the wind `efficiency' ($4L_{\rm mech} a^2/L_XR_s^2$) for the flow shown in Fig.~\ref{fig:30}.  In general this efficiency is low ($\la 0.1$) for planets with Neptune masses and above, indicating our assumption of radiative equilibrium is a good one, though we note that the efficiency increases to slightly higher values, $\sim 0.15$, for high density, earth-like, planets.  Jupiter mass planets drive the least efficient winds, with efficiencies of $\sim 0.01$.  This increase in efficiency with decreasing mass is most likely caused by the decrease in the dynamical timescale for planets with lower mass, whereas the thermal timescale remains roughly constant.  Additionally given that the mass-loss rate (and hence mechanical luminosity) scales linearly with X-ray flux, this indicates that X-ray evaporation is close to radiative-equilibrium at all X-ray luminosities and separations and that unlike the EUV case there is no transition from radiation-limited to energy-limited.  Thus, our calculations assuming radiative-equilibrium are accurate for the parameter space considered, and X-ray evaporation should not be thought of as `energy-limited'. Furthermore, while we point out that for sensible choices of the efficiency value, the `energy-limited' formalism can be used to recover correct values of the mass-loss rate, and give similar scalings, we caution that this efficiency value is certainly not fixed over orders of magnitude in both planetary mass and radius, and it cannot take into account how the increased area due to the position of the sonic surface varies with the parameters. Therefore, we find that Eq.~\ref{eqn:elimited} is of limited usefulness and would advocate using more accurate estimates of the mass-loss rates. { Furthermore, comparing these efficiencies with the calculations of \citet{cecchi2006}, who calculate various heating efficiencies for fixed ionization fractions ($10^{-1}-10^{-6}$) and find efficiencies between $80\%$ and $2\%$ respectively, we see that in all cases the bulk of the flow is at a fairly low ionization level.  Calculating the ionization fraction directly from the ionization parameter in our calculations we find ionization fractions $\ga 10^{-2}$ in the transonic region of the flow, in agreement with the static calculations of \citet{cecchi2006}, further validating the ionization parameter approach used in this work.}

\subsubsection{Multi-dimensional problem}
\label{sec:3D}
Planetary evaporation is ultimately a 3D problem (e.g. \citealt{stone2009}), and comparisons with observed line profiles will require knowledge of the 3D structure of the flow (\citealt{murrayclay2009}), and the role of day vs night-side heating. The models in this paper represent formal upper limits for the mass-loss rates, as we have assumed spherical outflow over the full $4\pi$ angular area of the planet. Future 3D radiation-hydrodynamical models of planetary evaporation will be able to shed light on the 3D structure of these flows (Owen et al. in prep), which will allow comparisons with observed line profiles, as lines sensitive to the flow outside the Roche radius will certainly be affected the 3D nature of the problem \citep{koskinen2010a}. Furthermore, they will be able to locally test some of the approximations (e.g. radiative equilibrium), as well as investigate the interaction between EUV and X-ray heating in the 3D setting. In this work we have assumed that the transition between X-ray driven and EUV driven occurs at a fixed point in time across the entire planetary wind, however, it is possible that parts of the planet could be undergoing EUV driven evaporation while other parts are X-ray driven.  Furthermore, a full 3D model will allow the assessment of whether non-axisymetric mass-loss can play a role in the orbital evolution of the planet, particularly for those undergoing significant fractional mass-loss (Neptune mass planets) as investigated by \citet{boue2012}.

\section{Conclusions}
\label{sec:conc}
In this work we have investigated the role of both X-ray and EUV heating in evaporating a close-in planet's atmosphere, using 1D hydrodynamic models.  We have considered how the mass-loss rates, and the driving mechanism, vary within the observable parameter space including: planet mass, planet density, separation, stellar mass, metallicity of the planet atmosphere and X-ray/EUV luminosity.  Then under the assumption that the EUV and X-ray luminosities follow an identical evolution in time, with the same initial luminosities, we have calculated the mass evolution of evaporating planets that evolve at constant density. Our main conclusions are as follows:
\begin{enumerate}
\item We identify two separate cases of planetary evaporation: X-ray driven, where the X-ray flow undergoes a sonic transition, and EUV driven, where the X-ray heated portion of the flow remains sub-sonic. In general X-ray evaporation occurs at high X-ray luminosities, low planetary densities, high planetary masses and small separations. Whereas EUV evaporation dominates at low X-ray luminosities, high planetary densities, low planetary masses and large separations.

\item We find that at separations of $<$0.1~AU, most planets will be evaporating hydrodynamically rather than through a ballistic mass-loss process. However, at separations greater than $\sim 0.1-0.5$~AU dense, Jupiter mass planets, may not be able to evaporate hydrodynamically. 

\item In the case of X-ray driven evaporation, the flow is close to radiative equilibrium and energy loss from the flow is dominated by line cooling.  The mass-loss rates scale linearly with X-ray luminosity and inversely with the separation squared ($\dot{m} \propto L_X/a^2$), but no exact scalings with planetary mass and planetary radii exist.

\item We find that as the high-energy luminosity falls over time, the evaporative flow may undergo a transition from X-ray driven at early times to EUV driven at late times.  This transition occurs at lower X-ray luminosities (and hence later times) for planets with at smaller separations.

\item Considering the evolution of Jupiter and Neptune mass planets, we find it is unlikely that an initially Jupiter mass planet can be completely evaporated within Gyr timescales, only losing a few percent of its original mass. However, we find that Neptune mass planets are much more susceptible to complete evaporation.  Within 3~Gyr Neptune mass planets at densities of $\la$0.3~g~cm$^{-3}$ are completely evaporated at a separation of 0.025~AU, while Neptune mass planets at densities of $\la$1 g cm$^{-3}$ are completely evaporated at a separation of 0.0125~AU. 

\item Using our evolutionary models we infer that at late times ($\sim$Gyrs), planets at separations $\la$0.025AU will still be undergoing X-ray driven evaporation, whereas at larger separations the evaporation will generally be in the EUV driven regime.
\end{enumerate}

\section*{Acknowledgments}
We thank the anonymous reviewer for comments which helped improve the paper. AJ is supported by an STFC Postgraduate Studentship. We thank Barbara Ercolano, Cathie Clarke, Norm Murray and Yanqin Wu for productive discussions. JO is grateful to hospitality from the IoA, Cambridge where this work began.

\appendix

\section{General X-ray flow solution}\label{app:gen}

In the case of a general X-ray heated flow, where the gas temperature is specified by a monotonic function of the ionization parameter ($T=f(\xi)$), we can follow a similar analysis to that given in Section~3, which gives an equation for the critical sub-to-super-sonic transition provided ${\rm d}\log f/{\rm d} \log \xi <1$ ($\alpha<1$ as in the illustrative case shown in Section~3.1) as
\begin{eqnarray}
u_{\rm s}^2&=&\frac{R_s}{2}\left.\frac{\partial \Psi_{\rm eff}}{\partial r}\right|_{r=R_S},  \\
f(\xi_s)\left(1-\left.\frac{{\rm d}\log f}{{\rm d} \log \xi}\right|_{\xi=\xi_s}\right) & = & \frac{R_s}{2}\left.\frac{\partial \Psi_{\rm eff}}{\partial r}\right|_{r=R_S},\label{eqn:xis}
\end{eqnarray}
where Eq.~\ref{eqn:xis} can be solved for the ionization parameter at the sonic surface ($\xi_s$). Using these conditions to specify the ionization parameter at a chosen sonic surface, as before we can obtain a transcendental function for the solution of $\xi(r)$, by applying mass conservation and using the Bernoulli potential. Thus the stream-function for the streamline connecting the planet and star is
\begin{equation}
\begin{split}
\frac{1}{2}\left(\frac{\xi}{\xi_s}\right)^2\left(\frac{r_s}{r}\right)^4u_s^2+& \frac{k_B}{\mu}\left[f(\xi)-\int^\xi\frac{f(\xi')}{\xi'}{\rm d}\xi'\right]\\+\Psi_{\rm eff}(r)= & H(R_s),
\end{split}
\label{eqn:bern2}
\end{equation} 
where $H(R_s)$ is the Bernoulli constant for a flow with a sonic transition at that given radius or
\begin{equation}
\begin{split}
H(R_s)=\frac{1}{2}\frac{R_s}{2}\left.\frac{\partial \Psi_{\rm eff}}{\partial r}\right|_{r=R_S}+&\frac{k_B}{\mu}\left[f(\xi_s)-\int^{\xi_s}\frac{f(\xi)}{\xi}{\rm d}\xi\right]\\ &+\Psi_{\rm eff}(R_s).
\end{split}
\end{equation}
Eq.~\ref{eqn:bern2} can now be solved for a family of solutions dependant on $R_s$ for a specified $T=f(\xi)$.

\bibliographystyle{mn2e}
\bibliography{JOAJrefs}

\label{lastpage}

\end{document}